\documentclass[12pt, letterpaper]{extarticle}

\usepackage{subcaption}
\usepackage{fullpage}
\usepackage[switch]{lineno}
\usepackage{amsmath}
\usepackage{amssymb}
\usepackage{rotating}
\usepackage{array}
\usepackage{mathtools}
\usepackage[ruled]{algorithm2e}
\usepackage{algorithmic}
\usepackage{bm}
\usepackage{breqn}
\usepackage{comment}
\usepackage{enumitem}
\usepackage{graphics}
\usepackage{graphicx}
\usepackage{latexsym}
\usepackage{mathrsfs}
\usepackage{morefloats}
\usepackage{nicefrac}
\usepackage{authblk}
\usepackage{pifont}
\usepackage{nameref}
\usepackage{booktabs}

\usepackage[hyphens]{url}
\usepackage{hyperref}
\hypersetup{colorlinks=false,breaklinks=true}

\title{On-Screen Inertia: Persistent Racial and Gender Disparities in Hollywood Film (1900-2024)}

\author[1]{Hazem Ibrahim}
\author[1]{Talal Rahwan}
\author[1,*]{Yasir Zaki}
\author[1,*]{Minsu Park}

\affil[1]{\normalsize New York University Abu Dhabi, UAE.}
\affil[*]{\footnotesize Correspondence author. E-mail: \{yasir.zaki, minsu.park\}@nyu.edu}
\date{}

\begin{document} 

\maketitle 

\begin{abstract}
\noindent 
Hollywood has diversified its casts. Whether this has translated into structural change in how those actors are positioned within narratives remains largely unexamined. Drawing on 76,815 U.S. English-language films (1900-2024) and over 3.1 million cast and crew entries, we move beyond headcounts to examine long-term inclusion trends through network centrality, occupational stereotypes, crew-to-cast diversity pathways, and financial outcomes. We find evidence of what we term on-screen inertia. While the raw inclusion of women and racial minorities has increased modestly, White actors have become more overrepresented relative to the U.S. Census in recent decades, not less. Within the visibility layer, women face a consistent longevity penalty with significantly shorter careers than men, and visual depictions framing men as dominant and women as sensual have remained stable since the 1950s. Structurally, White actors retain disproportionate network centrality; women achieve parity in centrality and lead billing yet cluster in secondary co-lead roles; and occupational stereotypes anchoring racial and gender groups to specific labor categories persist largely unchanged across the pre- and post-2000 periods. Crew diversity associates with cast inclusion only along matching demographic lines (i.e., racial with racial, 
gender with gender) and does not extend to narrative centrality, revealing a structural ceiling on hiring-based interventions. Critically, we find no consistent market penalty for diversity across decades of box office returns and audience ratings, eliminating the primary rationalization for these practices. Together, these findings demonstrate that Hollywood’s representational inequalities are not a rational market response, but are an institutionally sustained choice.

\end{abstract}

\clearpage

\section*{Significance Statement}
\noindent
Hollywood is widely seen as having grown more diverse, and measures of who appears on screen document that trend. But visibility alone cannot reveal whether diversity is structural or superficial. Analyzing 76,815 U.S. films (1900-2024) and 3.1 million cast and crew entries, we move beyond presence to measure narrative centrality, occupational stereotyping, visual portrayal, and financial outcomes. We find on-screen inertia: although visibility rose modestly, the hierarchies determining who anchors stories, what labor characters perform, and how they are visually portrayed barely moved. More diverse crews raise inclusion only within the same demographic dimension, neither crossing other dimensions nor extending to narrative centrality. Diverse films carry no box-office penalty, showing these disparities reflect an institutional choice, not a market necessity.

\section*{Introduction}
The stories told on screen shape how societies understand themselves. For more than a century, Hollywood has functioned not merely as entertainment, but as a cultural institution that reflects and reinforces dominant narratives of race, gender, and identity~\cite{cutting2016narrative, schweinitz2010stereotypes, gundlach1947movies, kumar2022gender, hedley1994presentation, baker1991role, gerbner2002growing}. Portrayals of characters, the roles they inhabit, and the narrative agency they are afforded both mirror and reproduce prevailing social hierarchies~\cite{ward2023media, tukachinsky2015documenting, erigha2015race, nwonka2020race, karniouchina2023women}. In this way, media representation operates as a mechanism through which cultural scripts are maintained, shaping audience perceptions and contributing to broader patterns of inequality.

Persistent inequalities in Hollywood have been consistently documented across both scholarly and public discourse. Movements such as \textit{\#OscarsSoWhite}~\cite{borum2018oscars, molina2018oscarssowhite} highlighted entrenched disparities, drawing attention to the ways in which representational hierarchies are embedded in the industry's structures and practices rather than isolated anomalies. Empirical research further demonstrates the systemic nature of these inequalities: women, racial minorities, and other marginalized groups remain underrepresented both on screen and in creative positions behind the scenes~\cite{smith2017inclusion, ramon2023hollywood, tukachinsky2015documenting, mastro2009racial, mastro2009effects, mastro2005latino}. These disparities are not simply a matter of numbers but reflect deeper institutional mechanisms that constrain access to resources, shape hiring and casting decisions, and ultimately reproduce social hierarchies through media production~\cite{erigha2015race, kuppuswamy2020testing, topaz2022race, karniouchina2023women}. Yet despite this extensive documentation, the dominant mode of evidence has remained headcount-based: scholars have measured \textit{who} appears on screen, but rarely \textit{where} those actors stand within the narrative structures they inhabit.

Recent computational advances have begun to push beyond this limitation. Bamman et al.~\cite{bamman2024measuring} conducted one of the first large-scale computational analyses of Hollywood films, introducing \textit{facetime}, the measured screen presence of demographic groups, as a benchmark for visibility, and establishing that facetime for marginalized groups has increased modestly over time. This work demonstrates the promise of computational methods for studying representation at scale. Yet its focus remains primarily on inclusion and screen presence: a \textit{visibility layer} of representation that, while necessary, captures only the surface of a deeper architecture. Complementary evidence from literary analysis underscores this limitation. Stuhler~\cite{stuhler2024agency} demonstrates that even when female characters are present in fiction, they are systematically assigned passive grammatical roles as objects of action rather than agents, revealing a structural agency gap that raw presence metrics cannot detect. While Stuhler's analysis operates at the level of literary syntax, it establishes a critical principle: visibility and structural positioning are distinct dimensions of representation, and progress on one does not guarantee progress on the other. Whether this agency gap extends to the structural positioning of characters within cinematic narrative networks, such as who occupies central roles, what labor they perform, and how they are visually framed, remains largely unexamined~\cite{topaz2022race, lauzen2020celluloid, smith2010gender}.

To address this gap, we propose that media representation must be evaluated across three distinct but interacting layers. The first is the \textit{visibility layer}: raw demographic inclusion, physical presence, and the portrayal standards (including age and visual framing) that condition whose visibility is permitted and on what terms. The second is the \textit{structural layer}: how characters are positioned within the narrative's social architecture, captured here through scene co-appearance network centrality, assigned narrative prominence, and the occupational stereotypes that define what labor marginalized actors perform on screen. The third is the \textit{institutional ecosystem layer}: the behind-the-scenes gatekeeping mechanisms---crew composition, director identity, and market incentives---that govern who gets cast and how centrally they are positioned. Together, these three layers allow us to distinguish between surface-level progress and structural change, and to identify precisely where Hollywood's representational hierarchies have yielded and where they have held firm.

Drawing on this framework, we assemble a dataset of 76,815 U.S. English-language films (1900-2024), comprising over 3.1 million cast and crew entries, supplemented with posters of 11,951 films for visual portrayal analyses~\cite{aldahoul2024inclusive} and scene-level co-appearance data for 3,265 films from the Amazon X-Ray Dataset~\cite{shrestha2026scene} for network-based analyses of narrative centrality (see Methods for details on dataset construction and analytical methods). Across all three layers, our analyses reveal a striking pattern of \textit{on-screen inertia}: while surface-level visibility has improved modestly over more than a century of American film, the structural hierarchies governing narrative prominence, occupational portrayal, and the distribution of agency have remained remarkably resistant to change. This resistance is inconsistent with a market-based rationale, pointing instead to institutionally sustained gatekeeping within the industry itself~\cite{smith2017inclusion, erigha2015race, kuppuswamy2020testing, karniouchina2023women}.

\section*{Results}
\label{sec:results}
\subsection*{The Conditionality of Visibility}

\begin{figure}
    \centering
    \includegraphics[width = \textwidth]{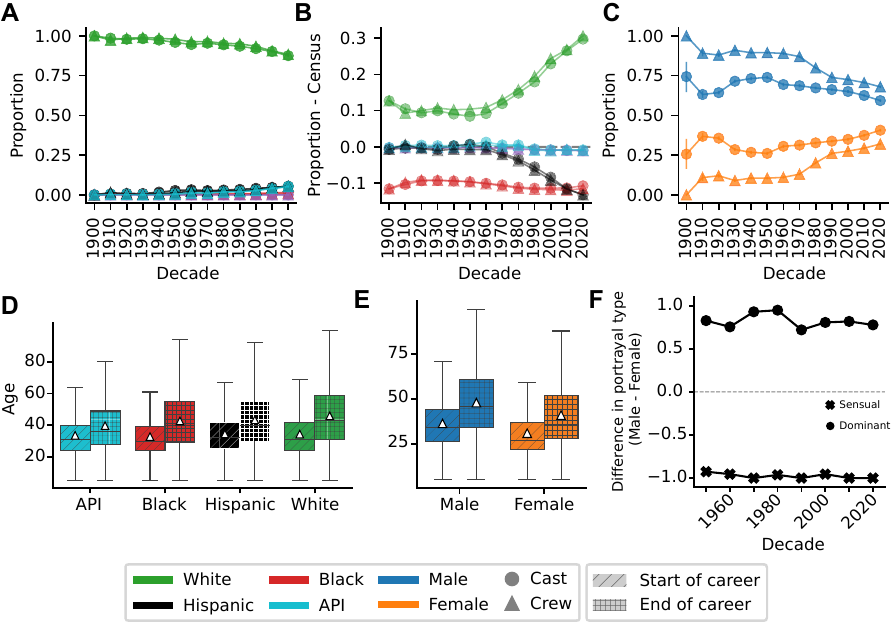}
    \caption{\textbf{The conditionality of visibility.} (\textbf{A}) The proportion of cast and crew positions occupied by a given racial group over time. (\textbf{B}) The proportion of cast and crew positions occupied by a given racial group relative to their census proportion in a given decade. (\textbf{C}) The proportion of cast and crew positions occupied by males and females over time. (\textbf{D}) Distribution of career entry and exit ages across racial groups. (\textbf{E}) Distribution of career entry and exit ages across gender groups. (\textbf{F}) The male-female difference in dominant and sensual portrayal types by decade, computed as the proportion of male depictions minus the proportion of female depictions in each category.}
    \label{fig:chapter_1}
\end{figure}

Prior work has documented modest increases in the absolute representation of racial minorities and women in cast compositions~\cite{bamman2024measuring, ramon2023hollywood, hunt2014hollywood}; we confirm this pattern and extend it to crews (Fig.~\ref{fig:chapter_1}A; see Methods for details on demographic attribute inference). White individuals nonetheless consistently held the overwhelming majority of cast and crew positions---exceeding 90\% from 1900 to 2010 and still comprising 88\% (cast) and 89\% (crew) of roles in the 2020s, yet these modest absolute declines do not match the rate of demographic change in the U.S. population. Benchmarked against U.S. Census population shares per decade (Fig.~\ref{fig:chapter_1}B), we find that White actors have become \textit{more} overrepresented in recent decades, not less: Hispanic representation has declined steadily since the 1960s relative to population share; Black individuals have remained persistently underrepresented across both centuries; and Asian/Pacific Islander actors were neither systematically over- nor under-represented. These racial representation patterns were remarkably consistent across cast and crew. Gender gaps follow a parallel but distinct pattern: while male individuals dominate both cast and crew roles, women are significantly more underrepresented in crew than in cast positions ($\chi^2(1, N = 439{,}892) = 1689$, $p < 0.001$; Fig.~\ref{fig:chapter_1}C), suggesting that behind-the-camera inclusion faces steeper structural barriers than on-screen presence. These patterns were consistently observed across the top three genres in our dataset (Drama, Comedy, and Thriller; Supplementary Fig.~11) and popularity tiers of films (Supplementary Fig.~2), confirming the robustness of our findings.

Beyond who appears on screen, disparities extend to how long they remain visible. White actors exit the industry at significantly older ages than all other racial groups ($\Delta_{\text{exit}} = 4.49$, $t = 23.73$, $p < 0.001$; White vs.\ all non-White), sustaining the longest observed careers (13.27 years), while Asian/Pacific Islander actors exhibit the shortest careers (7.31 years), followed by Hispanic (10.24 years) and Black actors (11.32 years; Fig.~\ref{fig:chapter_1}D). Entry age differences are more modest: White actors enter at slightly older ages on average ($\Delta_{\text{entry}} = 0.49$, $t = 3.22$, $p = 0.001$), though this difference is driven primarily by comparisons with API ($\Delta = 0.82$, $t = 4.02$, $p < 0.001$) and Black actors ($\Delta = 1.58$, $t = 2.56$, $p = 0.010$), with no significant difference relative to Hispanic actors ($\Delta = -0.06$, $t = -0.27$, $p = 0.789$). Gender disparities are more pronounced still: female actors enter Hollywood at significantly younger ages than male actors ($\Delta = 5.48$, $t = 54.1$, $p < 0.001$), exit earlier ($\Delta = 6.97$, $t = 55.1$, $p < 0.001$), and sustain significantly shorter careers overall ($\Delta = 1.77$, $t = 16.4$, $p < 0.001$; Figure~\ref{fig:chapter_1}E), a pattern consistent with the well-documented ``double standard'' of aging in entertainment~\cite{lauzen2005maintaining} that has persisted unchanged across the full span of our dataset (see Supplementary Note~1 for an extended analysis of cast member ages across racial and gender groups).

The terms on which visibility is granted reveal a further layer of conditionality. Figure~\ref{fig:chapter_1}F shows the difference in the proportion of dominant and sensual portrayals between male and female characters across decades. Dominant portrayals---characters depicted in commanding, authoritative, or imposing poses---are overwhelmingly male: on average, 92.1\% of individuals classified as dominant are men, with this proportion remaining stable across all eight decades (range: 88.0-97.4\%; linear trend $r = -0.10$, $p = 0.81$). Sensual portrayals---characters depicted in sexualized, alluring, or provocative poses---are almost exclusively female: on average, 99.0\% of individuals classified as sensual are women (range: 95.5-100.0\%; linear trend $r = -0.15$, $p = 0.72$). Neither trend shows a statistically significant change over the seven-decade observation window, indicating that the gendered visual grammar of movie posters has remained essentially frozen since the 1950s (see Methods for details on poster classification). To assess whether these patterns reflect genuine gendered characterization rather than generic visual-composition conventions, we drew on the face-ism framework of Archer et al.~\cite{archer1983face} as a comparison dimension: for each individual depicted on a poster, we classified body composition as full body, upper body, or face only, and computed the male-female difference in the proportion depicted face only by decade. Consistent with prior research documenting a moderate male skew in facial prominence, men are modestly more likely than women to be shown face only, but this difference is small throughout---peaking at roughly 2.5 percentage points in the 1950s, staying within about two percentage points across the intervening decades, and narrowing to essentially zero (parity) by the 2010s and 2020s (Supplementary Fig.~5). Unlike the dominant and sensual gaps (Figure~\ref{fig:chapter_1}F), which remain large and stable, the face-only gap effectively disappears. Hollywood has updated its compositional conventions (Supplementary Fig.~5), but it has not updated its characterization of gender.

Together, these findings suggest that surface-level progress masks deeper structural rigidities in how marginalized actors are positioned within narratives, which we examine next.

\if=0
\subsection*{Age analysis of racial and gender groups}
\fi

\subsection*{Structural Inertia}

\begin{figure}[htbp!]
    \centering
    \includegraphics[width = \linewidth]{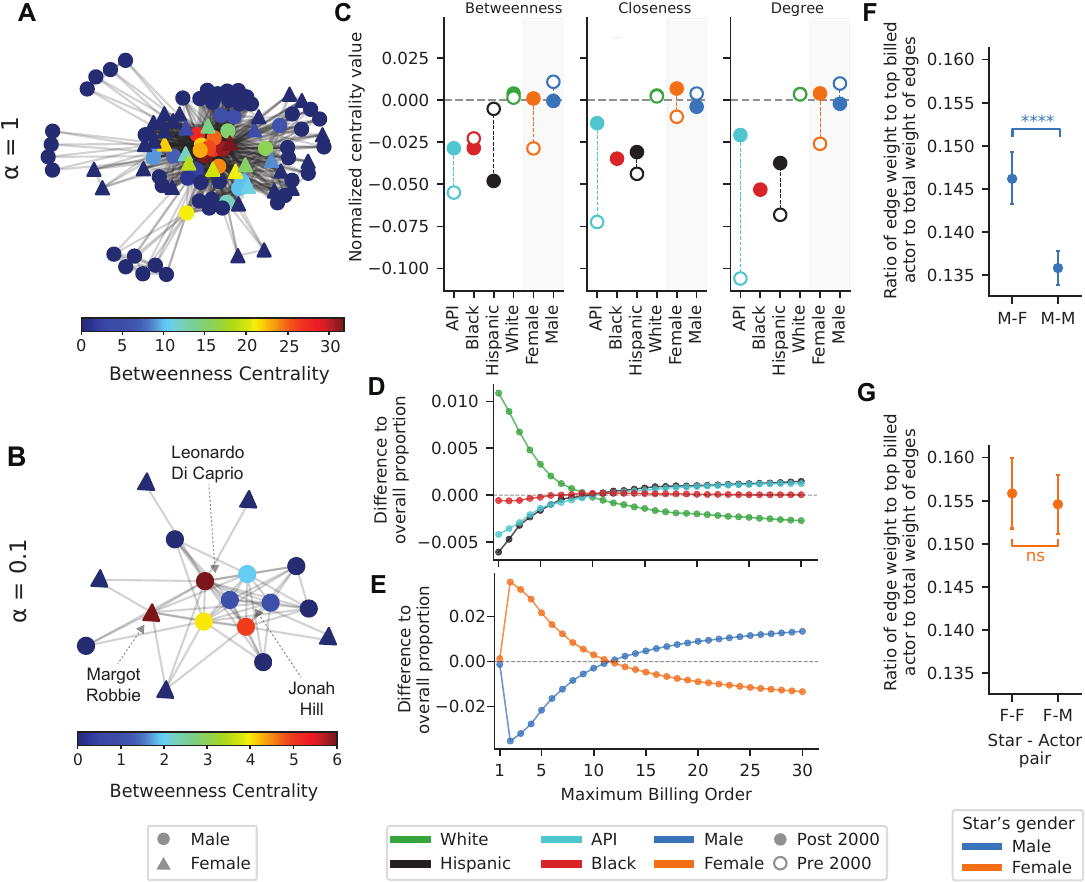}
    \caption{\textbf{Structural positioning in narrative networks.} (\textbf{A}, \textbf{B}) The actor co-appearance network for \textit{The Wolf of Wall Street} before and after applying the disparity filter algorithm ($\alpha = 0.1$), respectively. (\textbf{C}) Normalized centrality measures across betweenness, closeness, and degree for racial and gender groups in films produced before (open circles) and after (filled circles) 2000. (\textbf{D}) Deviation from expected representation for each racial group across billing positions, where positive (negative) values indicate overrepresentation (underrepresentation) at a given billing position. (\textbf{E}) The same analysis for gender groups. (\textbf{F}) Star ratio for male-led films, comparing female (M-F) and male (M-M) co-stars. (\textbf{G}) Star ratio for female-led films, comparing female (F-F) and male (F-M) co-stars.}
    \label{fig:chapter_2}
\end{figure}

Not all appearances are created equal. Even when marginalized actors are present on screen, the structural positions they occupy within narrative networks may differ systematically from those of their White and male counterparts. To examine this, we model each film as a weighted co-appearance network in which actors are nodes and edges reflect shared scene appearances, drawing on scene-level data from the augmented Amazon X-Ray Dataset~\cite{shrestha2026scene} for 3,265 films (see Methods for network construction and centrality normalization details). Figure~\ref{fig:chapter_2}A and B illustrate this approach using \textit{The Wolf of Wall Street} before and after applying the network backbone algorithm~\cite{serrano2009extracting}, which retains only the network's most structurally meaningful connections.

Across all three centrality measures, including betweenness, closeness, and degree, White actors occupy more central positions in narrative networks than expected, a pattern that has remained consistent across the pre- and post-2000 periods (Figure~\ref{fig:chapter_2}C). Because White actors numerically dominate the underlying casts, they inherently drive the baseline expected centrality. Consequently, their positive deviation from the mean is relatively small in magnitude. In stark contrast, all racial minority groups exhibit large, universally negative normalized centrality values. This reveals a severe structural penalty: when marginalized actors are included, they are systematically relegated to the periphery of the narrative network relative to expectation, regardless of the decade. Gender patterns are more nuanced: male and female actors show broadly comparable centrality overall, but female actors have moved toward greater centrality in films produced after 2000, while their male counterparts show a corresponding slight decline. These results are robust to the choice of disparity filter threshold $\alpha$ for the network backbone algorithm (Supplementary Tables~5-7), and are broadly consistent within the three most frequent genres in our dataset (Supplementary Fig.~12).

Billing order provides a complementary window into narrative hierarchy (Fig.~\ref{fig:chapter_2}D,E). White actors are overrepresented in top billing positions and underrepresented in peripheral roles, while racial minority actors are disproportionately concentrated in lower billing slots---with the exception of Black actors, who appear at roughly expected rates across positions but remain underrepresented in the highest-billed roles. The gender picture is more layered: female and male actors are equally likely to occupy the top billing slot relative to expectation, but female actors are substantially overrepresented at billing positions 2 through 11 and underrepresented at more peripheral positions beyond that range. This suggests that female characters frequently occupy prominent but secondary roles, such as co-leads or romantic interests, while male actors are more likely to appear either as the star or in minor background roles. These billing-order patterns hold when restricting the analysis to the top 5\% of films by popularity (Supplementary Fig.~4) and within the three most frequent genres in our dataset (Supplementary Fig.~13).

\begin{figure}[htbp!]
    \centering
    \includegraphics[width = \textwidth]{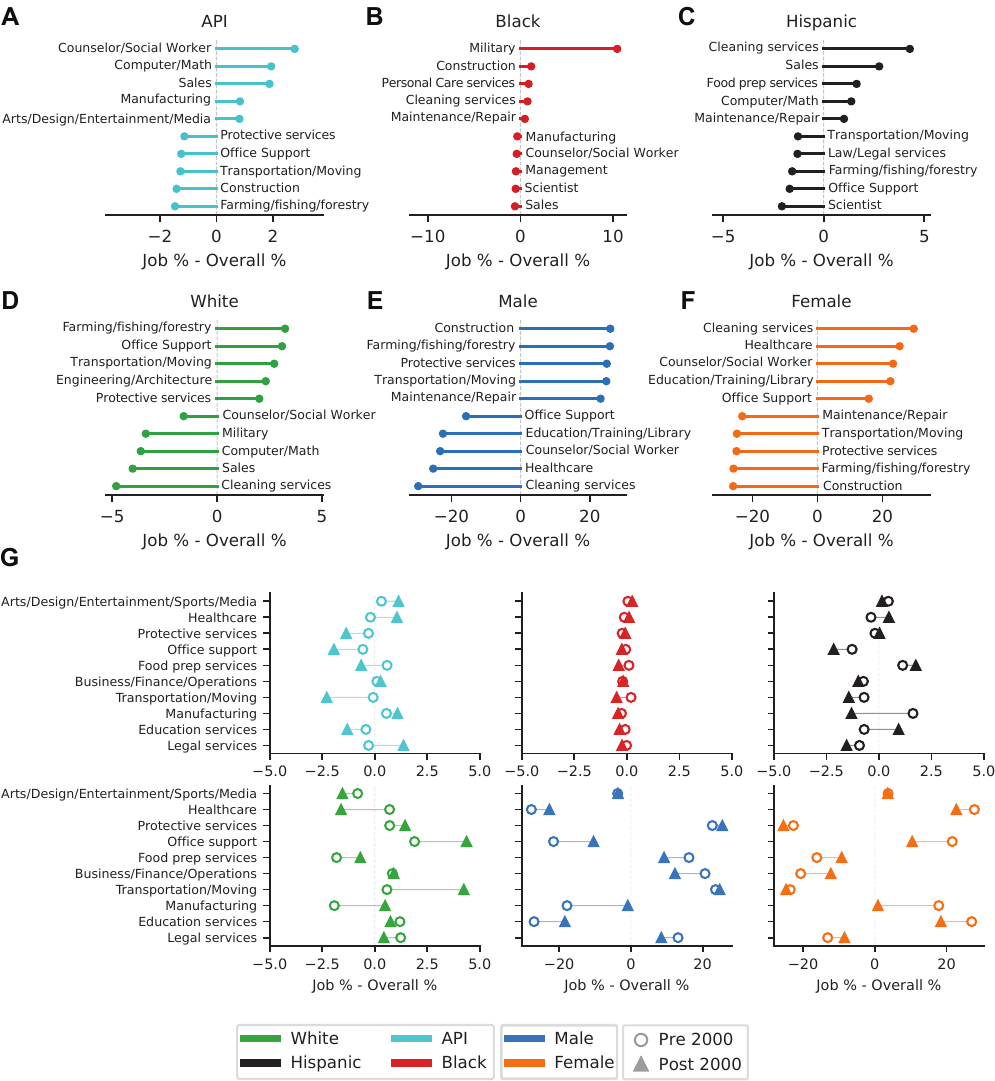}
    \caption{\textbf{Occupational stereotypes.} (\textbf{A}-\textbf{F}) The five most and five least representative occupational fields for each racial and gender group, relative to their overall frequency in the dataset. (\textbf{G}) Normalized representation within the ten most frequent occupational fields before (open circles) and after (filled triangles) the year 2000, for each racial and gender group.}
    \label{fig:chapter_3}
\end{figure}

To further characterize the gendered structure of narrative co-appearance, we compute the star ratio for each actor $A$ in a film with top-billed star $M$: 

\begin{equation*} 
    \text{Star ratio}(A) = \frac{w(E_{A \to M})}{\sum_{X \neq A} w(E_{A \to X})} 
\end{equation*} 

where $w(E_{A \to M})$ is the co-appearance edge weight between actor $A$ and the star, and the denominator is $A$'s total incident edge weight. A higher star ratio indicates that an actor's scene interactions are disproportionately concentrated around the star rather than distributed across the broader network. In films where the star is male, female actors show a significantly higher star ratio than their male counterparts ($t = 5.90$, $p < 0.001$; Fig.~\ref{fig:chapter_2}F), indicating that female actors in male-led films orient their scene interactions disproportionately around the male star rather than building independent narrative connections. This asymmetry disappears in female-led films, where male and female co-stars show comparable star ratios ($t = 0.99$, $p = 0.324$; Fig.~\ref{fig:chapter_2}G), suggesting that female stars do not generate the same gravitational pull on their co-stars' interactions. These patterns are consistent across $\alpha$ thresholds (Supplementary Table~8) and largely replicate within the three most frequent genres in our dataset (Supplementary Fig.~14).





Together, these network-based findings reveal a structural hierarchy that persists beneath the surface of apparent inclusion: White actors anchor narrative networks while minority actors populate their periphery, and women---even when present in prominent roles---remain relationally oriented toward male stars rather than serving as independent narrative hubs. Yet this network-based view, by design, focuses on named characters whose interactions define the narrative core. The peripheral positions---occupied by unnamed characters defined not by who they are but by what they do---carry implicit associations between identity and labor that are among the most visible and culturally resonant signals of who belongs where in society, yet have received comparatively little systematic attention. Mapping these occupational role labels to the O*NET taxonomy~\cite{onet2025} (see Methods for matching procedure), we find that the stereotypes governing this peripheral layer are both pervasive and remarkably stable across decades (Fig.~\ref{fig:chapter_3}).



Hispanic actors are disproportionately cast in cleaning services and food preparation roles, while appearing far less frequently as scientists or engineers (Fig.~\ref{fig:chapter_3}C). Asian/Pacific Islander actors are overrepresented in computer and mathematics fields (Fig.~\ref{fig:chapter_3}A). Black actors are overrepresented in military and construction roles while underrepresented in management and scientific occupations (Fig.~\ref{fig:chapter_3}B). Gender stereotypes are particularly pronounced: female actors cluster in healthcare, cleaning services, and education roles, while male actors dominate construction, protective services, and transportation (Fig.~\ref{fig:chapter_3}E,F). These patterns reflect real-world occupational segregation and reinforce cultural scripts about who belongs in which forms of labor.



To examine how these patterns have evolved over time, we compare representation across the ten most common occupational fields before and after the year 2000 (Fig.~\ref{fig:chapter_3}G). Across racial groups, representation within each occupational field remains remarkably stable, with minimal shifts between the two periods (the rank ordering of occupational over- and under-representation is significantly preserved across periods, Spearman $\rho = 0.45$, $p = .004$, with a mean absolute shift of only 1.41 percentage points and no significant change in the magnitude of deviations; paired $t = -1.39$, $p = .17$). Gender patterns show somewhat larger changes, with some movement toward the census baseline suggesting a slight compression in the extremes of over- and under-representation (the rank ordering of which fields are over- and under-represented for each gender is near-perfectly preserved, $\rho = 0.96$, $p < 0.001$, but the average magnitude of deviation from baseline has compressed significantly, from 23.3 to 17.5 percentage points; paired $t = 3.37$, $p = .003$). However, the structural boundaries of gendered labor on screen remain strikingly rigid: the occupational fields over-represented for men (e.g., Transportation, Protective Services) and women (e.g., Healthcare, Education Services) persist unchanged, indicating that while Hollywood may be moderately compressing the magnitude of its occupational biases, the underlying stereotypes themselves remain firmly entrenched.

\subsection*{Institutional Gatekeeping and Market Outcomes}
If structural hierarchies in narrative positioning are sustained from within the industry, the diversity of those making creative decisions---directors and crews---should predict who appears on screen and how centrally. Yet if gatekeeping operates selectively, these effects may extend to visibility while leaving narrative centrality intact. To examine this, we begin by examining the extent to which director identity predicts the composition of leading actors, quantifying director-actor demographic pairings through a normalized score that compares observed co-occurrence frequencies against a bootstrapped null model that shuffles billing order while holding film composition constant (see Methods for details). We then assess broader crew-cast diversity relationships and their downstream effects on narrative centrality.



\begin{figure}[htbp!]
    \centering
    \includegraphics[width = 0.95\textwidth]{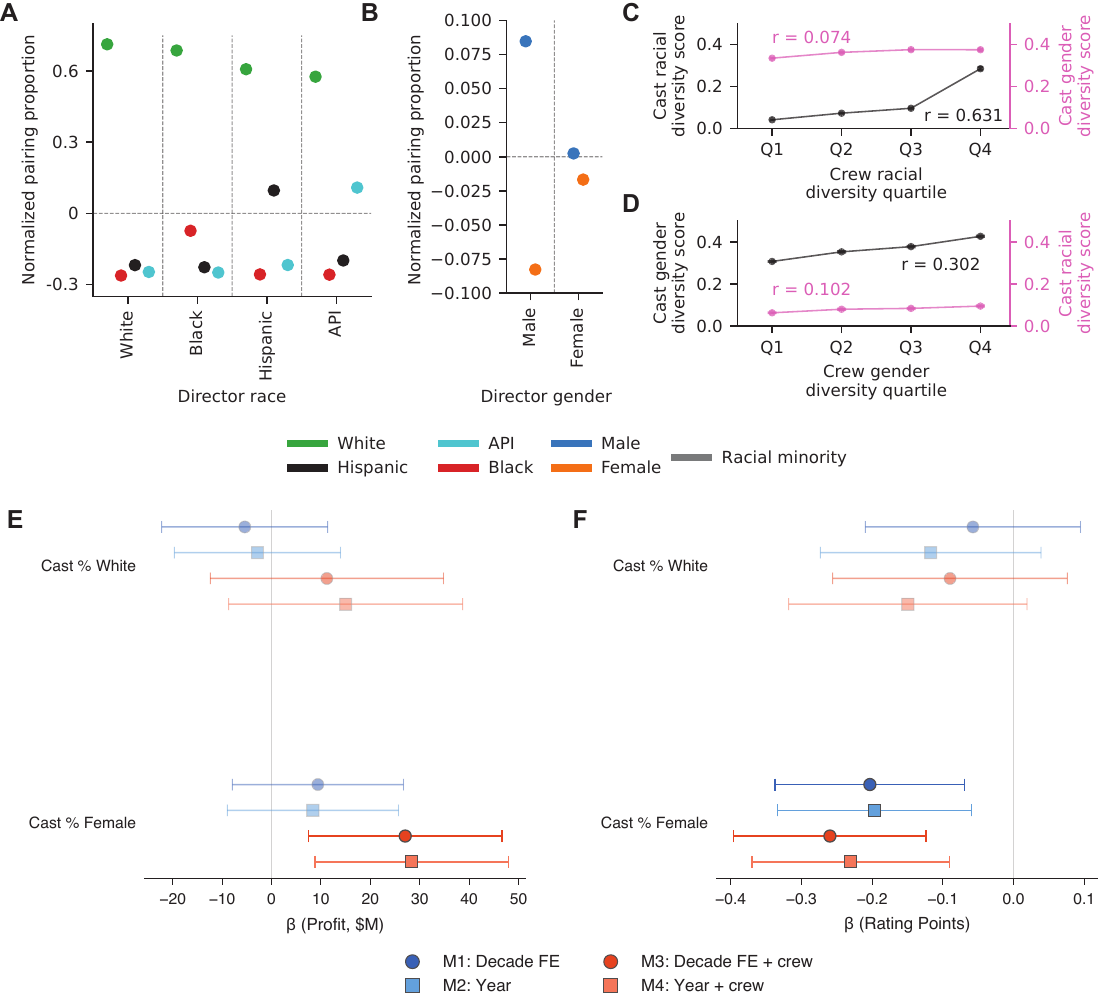}
\caption{\textbf{Institutional gatekeeping and market outcomes.} (\textbf{A}) Normalized likelihood of director-actor racial pairing, relative to a bootstrapped null model. Positive values indicate pairing frequencies exceeding chance. (\textbf{B}) Normalized likelihood of director-actor gender pairing. (\textbf{C}) Association between crew racial diversity quartile and cast racial diversity score (black, left axis) and cast gender diversity score (pink, right axis). (\textbf{D}) Association between crew gender diversity quartile and cast gender diversity score (black, left axis) and cast racial diversity score (pink, right axis). (\textbf{E, F}) Pooled OLS regression coefficients ($\beta$) for cast racial and gender diversity on film profit (\textbf{E}) and audience ratings (\textbf{F}) across four model specifications: M1 (decade fixed effects), M2 (year as continuous control), M3 (decade fixed effects with crew diversity controlled), and M4 (year with crew diversity controlled). All models include HC3 robust standard errors and control for log budget, log runtime, log cast and crew size, and primary genre. Color opacity indicates statistical significance: darker indicate $p < .05$; lighter indicate $p \geq .05$. Error bars represent 95\% confidence intervals.}
    \label{fig:chapter_4}
\end{figure}

White actors consistently dominate lead roles regardless of director race (Fig.~\ref{fig:chapter_4}A). However, among non-White directors, actors sharing the director's racial background appear more frequently in top-billed positions than would be expected by chance---a pattern most pronounced among Asian/Pacific Islander ($d = 0.11$) and Hispanic ($d = 0.10$) directors---suggesting a tendency toward racial alignment in casting decisions that partially counteracts the underrepresentation of same-race actors, but does not override, White dominance in lead roles (White actors retain the highest normalized pairing score across all director races, $d = 0.58$-$0.71$). This pattern holds across different billing-order cutoffs (Supplementary Table~9). A parallel pattern emerges for gender: male-directed films exhibit significant under-representation of female leads ($d = 0.16$, $p < 0.001$), whereas this gap narrows considerably in films directed by women ($d = 0.02$, $p < 0.001$; Fig.~\ref{fig:chapter_4}B), consistent across billing-order cutoffs (Supplementary Table~10). Both director-actor pairing patterns replicate within the three most frequent genres in our dataset (Supplementary Fig.~15A,B).


Extending this analysis to the broader crew level and incorporating cross-dimension comparisons, we find that diversity effects operate strictly within demographic dimensions rather than across them. We quantify racial diversity as the proportion of non-White individuals in the cast or crew, calculated as $1 - P(\text{race} = \text{White})$, and gender diversity as $1 - P(\text{gender} = \text{Male})$. Films with racially diverse crews show a strong positive association with racially diverse casts ($r = 0.631$, $p < 0.001$; Fig.~\ref{fig:chapter_4}C), but only a negligible association with gender-diverse casts ($r = 0.074$, $p < 0.001$), indicating that racial diversity behind the camera does not translate into greater gender diversity on screen. The pattern is symmetric: gender-diverse crews are moderately associated with gender-diverse casts ($r = 0.302$, $p < 0.001$; Fig.~\ref{fig:chapter_4}D) but show little relationship with racial cast diversity ($r = 0.102$, $p < 0.001$). Representational influence is thus dimension-specific, as diversity behind the camera shapes inclusion on screen only within, not across, demographic lines (see Supplementary Fig.~15C,D for robustness across the three most frequent genres in our dataset). Extending this analysis to narrative centrality, we find that while individual director identity can partially narrow centrality gaps for marginalized actors, these effects dissipate when considering overall crew diversity, revealing a ceiling effect on inclusion-based interventions that operates even within demographic dimensions (Supplementary Fig.~6 and Tables~11-16). 

Finally, we test whether these structural disparities reflect rational market responses to audience preferences, as the consumer discrimination hypothesis predicts~\cite{becker2010economics, kuppuswamy2020testing} and as the industry has frequently cited to justify exclusionary casting and hiring practices~\cite{bielby1994all, erigha2015race, hunt2014hollywood}. Across four pooled OLS specifications with HC3 robust standard errors, we estimate the associations between cast diversity and financial and audience outcomes, progressively controlling for crew composition to isolate on-screen effects from behind-the-scenes structural factors (Figs.~\ref{fig:chapter_4}E,F; see Supplementary Table~17 and Supplementary Fig.~7 for full model results). Cast racial composition shows no significant association with either profit or audience ratings across any model ($p = 0.21$-$0.74$ for profit; $p = 0.08$-$0.46$ for ratings), providing no support for a racial market penalty. For gender, the results depend critically on whether crew composition is controlled. Without crew controls (M1 and M2), Cast \% Female shows no significant association with profit. Once crew diversity is held constant (M3 and M4), Cast \% Female becomes significantly and positively associated with profit ($\beta \approx 27$-$28$, $p < .007$). This may reflect a suppression effect in which behind-the-scenes structure was masking a positive on-screen gender signal: films with more women on screen are associated with higher returns once crew composition is accounted for. This positive profit association holds within individual genre subsets and is robust to more granular genre controls (Supplementary Figs.~8 and 9). Cast \% Female shows a modest but consistent negative association with audience ratings across all models ($\beta \approx -0.20$ to $-0.26$, $p < .005$). Importantly, this rating penalty does not translate into a profit penalty and is concentrated in Action and Thriller (Supplementary Figs.~8 and 9), the two genres with the most male-skewed platform audiences. This pattern is consistent with the well-documented demographic skew of online rating platforms, where male-skewed user bases have been shown to systematically down-rate female-centric content~\cite{stroube2024status, aguiar2026bad}. The divergence between what platform raters signal and what the paying market does is itself evidence against consumer discrimination as an economic rationale. Taken together, these findings undermine the market-based rationalization for Hollywood's demographic disparities: on-screen diversity does not penalize films at the box office, and the financial disadvantages that do exist originate from within the industry rather than from consumer demand~\cite{kuppuswamy2020testing, karniouchina2023women}.

\section*{Discussion}
Across more than a century of American film, our analyses reveal a striking pattern of \textit{on-screen inertia}: surface-level visibility has improved modestly, but the structural hierarchies governing narrative prominence, occupational portrayal, and visual framing have remained remarkably resistant to change. This resistance is inconsistent with a purely market-based explanation and points instead toward factors operating within the industry itself. Our findings address a foundational tension in the study of media representation, namely the gap between visibility and structural positioning. Prior work established that the measured screen presence of marginalized groups has increased modestly over time~\cite{bamman2024measuring}, and complementary evidence from literary analysis demonstrated that even when female characters are present, they are systematically assigned passive roles as objects of action rather than agents~\cite{stuhler2024agency}. Our study bridges these two contributions by showing that the same gap between visibility and structural positioning governs cinematic narrative, where it has persisted largely unchanged across more than a century of Hollywood film.

The conditionality of visibility documented here has implications for how we understand the relationship between institutional change and cultural output. Hollywood has responded to external diversity pressures by increasing the numerical presence of marginalized actors, yet the terms of that presence, including the portrayal standards, career structures, and visual grammar through which it is delivered, have barely moved. The persistence of the longevity penalty for women and the frozen visual grammar of dominance and sensuality since the 1950s suggest that the industry has separated surface-level casting decisions from the deeper ideological conventions that govern how bodies are framed and careers are sustained. This divergence is itself a form of institutional inertia: organizations can update their demographic composition while leaving their cultural logic intact~\cite{erigha2015race, karniouchina2023women}.

At the structural layer, our network-based findings extend the agency gap identified in literary fiction~\cite{stuhler2024agency} to cinematic narrative. Female actors, even when present in prominent roles, remain relationally oriented toward male stars rather than serving as independent narrative hubs, raising the possibility that this agency gap is not unique to a particular medium but reflects a broader pattern in how gender is encoded across both literary and cinematic traditions. The resilience of occupational stereotypes adds a further dimension. Even the most peripheral and unnamed characters carry implicit associations between identity and labor that cultural scripts have maintained with minimal compression across pre- and post-2000 periods, and these background stereotypes may be precisely the ones that operate below the threshold of conscious audience attention, making them among the most resilient and the least examined in prior large-scale work~\cite{topaz2022race, lauzen2020celluloid}.

Perhaps the most consequential finding for policy is the ceiling effect on hiring-based interventions. Crew diversity predicts cast inclusion only along matching demographic lines: racially diverse crews hire more racially diverse actors and gender-diverse crews hire more women, but neither crosses demographic dimensions. Diversity initiatives thus remain siloed by identity, and a studio that increases crew gender diversity cannot claim it is simultaneously advancing racial inclusion, and vice versa. Even within these bounded effects, crew diversity does not shift narrative centrality. Current policy instruments may be addressing a symptom, who is on screen, without addressing the underlying condition, where those actors stand within the narrative and what agency they are afforded. Initiatives that stop at hiring may therefore produce visible diversity without structural equity, a distinction with direct consequences for how we design and evaluate representational interventions.

The market-based rationalization for Hollywood's demographic disparities does not hold up under empirical examination. Neither racial diversity nor female cast composition penalizes profit, suggesting that the industry's claim that diverse films are commercially unviable reflects a perceived rather than an empirical reality~\cite{kuppuswamy2020testing}. The concentration of the ratings penalty for female cast composition in Action and Thriller genres, where platform audiences are most male-skewed, further suggests that the signal studios receive from online ratings platforms is not representative of the broader paying market~\cite{stroube2024status, aguiar2026bad}. That studios have historically treated this platform signal as evidence of consumer discrimination may itself function as a mechanism of institutional gatekeeping: a rationalization that is self-reinforcing precisely because it is rarely tested directly. Our findings test it and find it wanting.

\subsection*{Limitations}
Several limitations bound these findings. Our demographic inferences rely on name-based algorithmic classification~\cite{golder2022methods, vanhelene2024inferring}, which introduces misclassification error that is not uniformly distributed: minority names tend to receive lower confidence scores due to underrepresentation in classifier training data, meaning our confidence threshold likely excludes minority individuals at disproportionately higher rates~\cite{lockhart2023name}, and our estimates of minority underrepresentation should be interpreted as conservative lower bounds. Our analyses are restricted to U.S.\ English-language films, leaving open whether on-screen inertia generalizes across global film industries. Network centrality and occupational coding, while powerful proxies, cannot capture dialogue content, screen time, or the subjective richness of character arcs, and the network analyses rely on a subset of 3,265 films from the augmented Amazon X-Ray dataset~\cite{shrestha2026scene} that may not fully represent the broader population. The visual portrayal analysis is based on film posters rather than on-screen content~\cite{aldahoul2024inclusive}, and while our market results provide strong evidence against a consistent financial penalty for diversity, they cannot rule out consumer discrimination in specific contexts or time periods, a question that experimental designs and platform-specific viewer data are better positioned to address. More broadly, our analyses are correlational rather than causal: while the consistency of these patterns across demographic dimensions, time periods, and outcome measures is difficult to reconcile with a purely market-based account, we cannot rule out unmeasured confounds, such as budget allocation or other unobserved production decisions, and our findings should not be read as identifying the specific decision-making mechanisms responsible for the disparities we document.

Taken together, our analyses reveal a fundamental tension between progress and persistence in Hollywood's approach to representation. Marginalized groups are increasingly visible on screen, but they remain constrained by structural hierarchies of narrative prominence, occupational stereotyping, rigid portrayal standards, and entrenched institutional practices that hiring-based interventions alone have proven insufficient to dismantle. The absence of a market-based rationale for these disparities indicates that on-screen inertia is not an economic inevitability but an institutionally sustained choice. The industry retains both the latitude and, our findings suggest, the financial incentive to change these patterns.

\section*{Data and Methods}

\subsection*{Data sources and study scope}
We constructed a century-scale dataset of Hollywood films using The Movie Database (TMDB) API, querying all U.S.\ English-language films released between 1900 and 2024 with the United States listed as a country of origin, yielding an initial dataset of 332{,}356 films (see Supplementary Table~1 for a breakdown by decade). For each film, we retrieved metadata (e.g., release year, popularity metrics) and full credits for cast and crew, including cast billing order and crew department and job titles, resulting in over 4.5 million individual entries prior to filtering. To contextualize representation patterns, we used decennial U.S.\ Census population shares as benchmarks when computing relative representation by decade. Occupational analyses relied on the O*NET taxonomy to map character occupation labels to standardized occupational fields~\cite{onet2025}.

\subsection*{Preprocessing and inclusion criteria}
We applied two filters to ensure analyses were based on sufficiently informative credits data. First, we removed films with fewer than five credited cast or crew members to reduce noise from incomplete or atypical entries common in early and low-budget productions. Second, we inferred perceived race and gender for each credited individual using name-based classifiers and removed entries for which classifier confidence was $\leq 0.5$. After filtering, our final analytic sample comprised 76{,}815 films and 3{,}118{,}706 cast and crew entries. Throughout the paper, we treat cast and crew entries as role-level observations, as an individual may appear in multiple films and multiple roles. Unless otherwise specified, statistics are computed over role-level observations, with decade-based aggregation for long-term trend analyses.

\subsection*{Demographic inference from names}
We inferred perceived race using NamePrism~\cite{ye2017nationality, ye2019secret} and perceived gender using Genderize~\cite{genderize_2019}, tools widely used in computational social science to estimate perceived race and gender from names~\cite{golder2022methods, vanhelene2024inferring, kempf2021partisan, luca2024evolution}. Both tools output a predicted category and an associated confidence score. For race, we aggregated NamePrism outputs into four broad groups: White, Black, Hispanic, and Asian/Pacific Islander (API). For gender, we used binary categories (Male/Female), consistent with the output of Genderize. In both cases, we discarded predictions with confidence $\leq 0.5$ (see Supplementary Table~2 and Supplementary Fig.~3 for the distribution of confidence scores and the share of entries passing different thresholds).

\subsection*{Visual portrayal classification on movie posters}
To characterize how actors are visually framed on promotional materials, we collected posters for U.S.\ English-language films from the TMDB API, yielding 11{,}951 posters spanning eight decades (1950s: $n = 1{,}574$; 1960s: $n = 1{,}703$; 1970s: $n = 1{,}645$; 1980s: $n = 1{,}610$; 1990s: $n = 1{,}531$; 2000s: $n = 1{,}556$; 2010s: $n = 1{,}527$; 2020s: $n = 745$).
 
We adopted the classification pipeline introduced in \cite{aldahoul2024inclusive}, which uses GPT-4 with Vision~\cite{openai2024gpt4v} to classify uncropped poster images along three dimensions: \emph{body composition} (full body, upper body, or face only), \emph{posture} (standing, sitting, or reclining), and \emph{portrayal type} (dominant, sensual, or submissive). The model was prompted on each poster image to describe the gender, posture, visible body parts, and portrayal context of every individual depicted. Portrayal categories were derived through a data-driven procedure: free-text descriptions were embedded using OpenAI's \texttt{text-embedding-3-large} model, clustered via $k$-means ($k = 20$), and then consolidated into coherent categories through manual inspection. Three categories, namely dominant, sensual, and submissive, were retained as the primary portrayal types based on cluster coherence, with each operationalized using a set of 20 semantically related keywords (10 adjective-adverb pairs) identified via FastText~\cite{fasttext} nearest-neighbor lookup in the embedding space (e.g., \emph{dominant}: commanding, imposing, imperious; \emph{sensual}: seductive, alluring, erotic; \emph{submissive}: docile, meek, obedient). An image was assigned to a portrayal category if its GPT-4 description contained any of the corresponding keywords; images could be assigned to multiple categories. The submissive category was retained in the classification pipeline but excluded from main text analyses due to insufficient observations for reliable decade-level estimation in U.S.\ English-language films.
 
Validation was performed by three independent human coders who annotated samples of 200 images each. GPT-4's portrayal classifications achieved an accuracy of 88.1\%, with inter-rater agreement of Krippendorff's $\alpha > 0.86$ and Cohen's $\kappa > 0.6$ between the majority human label and the model output. Gender classification achieved 99.5\% accuracy ($\kappa = 0.99$). Full validation details are provided in \cite{aldahoul2024inclusive}.
 
For each decade, we computed the proportion of dominant and sensual portrayals attributed to male versus female actors and assessed temporal stability using OLS regression of the male share of dominant portrayals and the female share of sensual portrayals on a linear decade index. As a comparison dimension to assess whether the dominant and sensual gaps reflect genuine characterization ideology rather than generic compositional conventions, we computed the face-only portrayal gap (male minus female proportion) across decades, following the face-ism framework of Archer et al.~\cite{archer1983face} (Supplementary Fig.~5).

\subsection*{Representation metrics}
For each decade and group $g$, we computed the share of 
cast or crew roles attributed to $g$ as
\begin{align*}
    P_{\text{film}}(g \mid d) = \frac{\#\,\text{roles in 
    decade } d \text{ attributed to } g}{\#\,\text{all 
    roles in decade } d},
\end{align*}
and relative representation by comparing to the 
corresponding U.S.\ Census benchmark 
$P_{\text{census}}(g \mid d)$ as
\begin{align*}
    \Delta(g \mid d) = P_{\text{film}}(g \mid d) - 
    P_{\text{census}}(g \mid d).
\end{align*}

\subsection*{Age and career-trajectory analysis}
To examine age disparities, we used actor birthdates and computed per-role ages as release year minus birth year, excluding entries with missing values (e.g., an actor with no recorded birthdate). We excluded actors whose most recent appearance was in 2020 or later, as their careers may still be ongoing and the dataset's end date would artificially truncate their observed career length. Age values were restricted to the 5-100 range to exclude implausible entries. For each demographic group, we summarized the distribution of per-role appearance ages and compared group means using independent two-sample $t$-tests.

Career entry was defined as the earliest release year in which an actor appears in our dataset, and career exit as the latest. Entry age, exit age, and career length (exit year minus entry year, plus one) were computed accordingly. Extended age analyses across racial and gender groups are reported in Supplementary Note~1.

\subsection*{Movie co-appearance networks and narrative centrality}
\paragraph{Network construction.}
To quantify narrative prominence, we modeled each film as a weighted undirected network in which nodes are actors and an edge between actors $i$ and $j$ indicates that the two co-appear in the same scene, with edge weight $w_{ij}$ equal to the number of scenes in which they co-appear. Scene-level character appearance data were obtained from the augmented Amazon X-Ray dataset~\cite{shrestha2026scene}, which provides curated scene breakdowns for 3{,}265 U.S.\ movies sourced from the Amazon Prime Video X-Ray, recording scene boundary timestamps and characters visible in each scene with IMDb identifiers. This linkage enabled us to merge scene co-appearance data with the demographic labels inferred from our TMDB-based pipeline by matching actor identities across the two databases. Unlike screenplay- or subtitle-based extraction methods, which are susceptible to character disambiguation errors and discrepancies between draft scripts and final films, the X-Ray data reflects actual on-screen content, providing a more reliable foundation for network construction.

\paragraph{Backbone extraction via disparity filter.}
To reduce spurious connections and more accurately capture the structure of each movie network, we applied the disparity filter of Serrano et al.~\cite{serrano2009extracting}. For a given node $i$ with degree $k_i$ and normalized incident edge weights $p_{ij} = w_{ij}/\sum_{l} w_{il}$, the disparity filter tests whether an edge weight is unexpectedly large under a null model in which $i$’s total strength is uniformly distributed across its $k_i$ edges. We retained edges whose significance level is below a fixed threshold $\alpha$ (note that we used $\alpha=0.1$ for the illustrative networks in Fig.~\ref{fig:chapter_2}A,B). Centrality results are robust to the choice of $\alpha$; betweenness, closeness, and degree centrality differences across a range of $\alpha$ values are reported in Supplementary Tables~5-7.

\paragraph{Centrality measures.}
On each filtered backbone, we computed degree, closeness, and betweenness centrality for every actor. To compare centrality across movies of different sizes, we normalized each actor's centrality by subtracting the mean centrality within the corresponding bucket defined by centrality measure, cast-size octile (eight quantile-based bins of cast size), and release year, yielding a mean-centered deviation score in the original units of each centrality measure. 

\paragraph{Star ratio.}
To quantify how strongly a character's interactions are oriented around the top-billed actor (billing order 1), we computed, for each actor $A$ in a film with star actor $M$,
\begin{align*}
    \text{Star ratio}(A) = \frac{w_{AM}}{\sum_{X \neq A} w_{AX}},
\end{align*}
where $w_{AM}$ is the co-appearance edge weight between $A$ and the star actor and the denominator is $A$’s total incident weight.

\subsection*{Billing-order analysis and randomized null model}
To assess whether demographic groups are over- or under-represented at specific billing positions, we computed, for each group $g$ and billing position $N$, the fraction of group members appearing at $N$ within each film. We then constructed a bootstrapped null model by randomly permuting billing order within each film while holding the set of credited actors and their demographic labels fixed. Repeating this permutation procedure 1{,}000 times yielded an expected distribution for each $(g, N)$, from which we computed a normalized difference between the observed and expected frequencies, where positive values indicate over-representation and negative values indicate under-representation at a given billing position.

\subsection*{Occupational-role mapping using O*NET}
We identified occupational roles by matching character role descriptors in cast credits (e.g., ``Nurse'', ``Detective'', ``Waiter'') to official occupation titles and synonyms in O*NET, a comprehensive database that categorizes occupations across the U.S.\ economy~\cite{onet2025}. Matched roles were mapped to O*NET's broader occupational fields, allowing analysis across standardized employment categories (see Supplementary Table~4 for the number of matched roles by occupational field and decade). For each demographic group $g$ and occupational field $o$, we computed a representation score as the deviation between $g$'s share of roles within $o$ and $g$'s overall share across all occupationally-matched roles in our film sample. We used this within-sample baseline, rather than external labor force statistics, as consistently classified race- and gender-specific employment data are not available at the required occupational granularity across our full study period. 

To examine temporal stability, we repeated this analysis for films released before versus after 2000, focusing on the ten most frequent occupational fields. We assessed stability using Spearman rank correlations between pre- and post-2000 normalized representation scores and tested for compression in the magnitude of deviations using paired $t$-tests on absolute representation scores.

\subsection*{Crew-cast diversity relationships}
\paragraph{Diversity indices.}
We operationalized racial diversity of a film's cast or crew as $1 - P(\text{race} = \text{White})$ and gender diversity as $1 - P(\text{gender} = \text{Male})$, computed at the film level.

\paragraph{Director-actor pairing analysis.}
For director-actor demographic pairing, we computed the observed frequency of each director group paired with each actor group among the top three billed actors. We compared these frequencies to a bootstrapped null model that preserves the set of actors and their demographics in each film but randomizes billing order, yielding a normalized observed-minus-expected pairing score. To verify that the results are not an artifact of the top-three cutoff, we repeated this analysis using the top 3, 5, 10, 20, and 30 billed actors; the pairing patterns are stable across all cutoffs (Supplementary Tables~9 and 10 for racial and gender pairing, respectively).


\paragraph{Correlations with on-screen outcomes.}
We assessed associations between crew diversity and cast diversity using Pearson correlation coefficients computed on per-film raw diversity scores, with quartile binning used only for visualization. To test whether behind-the-camera diversity predicts narrative centrality, we computed, for each film, the betweenness-centrality gap between marginalized and majority actor groups (non-White minus White for the racial dimension; female minus male for the gender dimension), using the raw co-appearance-graph betweenness, and correlated these film-level gaps with the film's crew racial and gender diversity scores, both within and across demographic dimensions (Supplementary Fig.~6C,D). Separately, we compared actor centrality across all three measures (degree, closeness, betweenness) between films grouped by director racial and gender majority, across a range of top-$N$ billing cutoffs (Supplementary Figs.~6A,B and Supplementary Tables~11-16).

\subsection*{Financial and audience outcome analysis}
To test whether cast diversity predicts financial or audience outcomes, we retrieved budget and revenue data from the TMDB movie details API for all films in our dataset. We computed profit as revenue minus budget (in millions of USD) and used TMDB user ratings (vote average) as a measure of audience evaluation, restricting the ratings sample to films with at least 50 votes to reduce noise from sparsely rated entries. After applying the full set of model controls, the profit sample comprised 8{,}298 films with non-missing budget, revenue, and control variables; the ratings sample comprised 8{,}622 films meeting the vote threshold with non-missing controls. Decade-by-decade replications using the full available samples (9{,}516 films for profit; 19{,}133 films for ratings) are reported in Supplementary Fig.~10.

We estimated four pooled OLS model specifications, all 
with HC3 heteroscedasticity-consistent robust standard 
errors:
\begin{itemize}[nosep, label=--]
    \item \textbf{M1} (decade fixed effects): $Y \sim \text{diversity variable} + \log(\text{budget}) + \log(\text{runtime}) + \log(\text{cast size}) + \log(\text{crew size}) + \text{genre} + \text{cross-diversity control} + C(\text{decade})$
    \item \textbf{M2} (year as continuous control): same as M1 but replacing decade fixed effects with year as a continuous predictor.
    \item \textbf{M3} (decade FE + crew controls): estimates cast diversity effects only, adding crew \% White and crew \% Female as controls to isolate on-screen effects from behind-the-scenes structural factors.
    \item \textbf{M4} (year + crew controls): same as M3 but with year as a continuous predictor.
\end{itemize}

\noindent where $Y$ is either profit (in millions of USD) or TMDB vote average. Cast racial composition was operationalized as the proportion of White cast members and cast gender composition as the proportion of female cast members. In M1 and M2, each diversity variable was estimated in a separate regression with a same-dimension cross-diversity control (e.g., when estimating Cast \% White, Cast \% Female was included as a control, and vice versa). In M3 and M4, both crew diversity measures were included simultaneously as controls when estimating cast effects, isolating the on-screen diversity signal from crew composition. Primary genre was included as a categorical fixed effect, defined as the first listed genre in a film's TMDB metadata. Full model results, including crew diversity coefficients, are reported in Supplementary Table~17 and Supplementary Fig.~7.

As a robustness check, we conducted genre-stratified analyses in two specifications: (A) estimating M3 and M4 separately within each of the seven most frequent genres in our sample (Action, Adventure, Comedy, Drama, Horror, Romance, Thriller), and (B) replacing the single primary-genre fixed effect with binary indicator variables for each of these seven genres, allowing films tagged with multiple genres to contribute to multiple categories simultaneously and addressing the limitation that primary genre captures only the first listed genre tag. Results are reported in Supplementary Figs.~8 and 9.

\subsection*{Statistical testing and reporting}
We report two-sided hypothesis tests throughout. For mean comparisons, we use independent two-sample $t$-tests; for categorical comparisons, we use $\chi^2$ tests. Significance thresholds are $p < 0.05$, and exact $p$-values are reported when available.

\bibliographystyle{naturemag}
\bibliography{sample}

\begin{thebibliography}{10}
\expandafter\ifx\csname url\endcsname\relax
  \def\url#1{\texttt{#1}}\fi
\expandafter\ifx\csname urlprefix\endcsname\relax\def\urlprefix{URL }\fi
\providecommand{\bibinfo}[2]{#2}
\providecommand{\eprint}[2][]{\url{#2}}

\bibitem{cutting2016narrative}
\bibinfo{author}{Cutting, J.~E.}
\newblock \bibinfo{title}{Narrative theory and the dynamics of popular movies}.
\newblock \emph{\bibinfo{journal}{Psychonomic bulletin \& review}}
  \textbf{\bibinfo{volume}{23}}, \bibinfo{pages}{1713--1743}
  (\bibinfo{year}{2016}).

\bibitem{schweinitz2010stereotypes}
\bibinfo{author}{Schweinitz, J.}
\newblock \bibinfo{title}{Stereotypes and the narratological analysis of film
  characters}.
\newblock In \emph{\bibinfo{booktitle}{Characters in fictional worlds:
  Understanding imaginary beings in literature, film, and other media}},
  \bibinfo{pages}{276--289} (\bibinfo{publisher}{De Gruyter},
  \bibinfo{year}{2010}).

\bibitem{gundlach1947movies}
\bibinfo{author}{Gundlach, R.~H.}
\newblock \bibinfo{title}{The movies: Stereotypes or realities?}
\newblock \emph{\bibinfo{journal}{Journal of Social Issues}}
  \textbf{\bibinfo{volume}{3}}, \bibinfo{pages}{26--32} (\bibinfo{year}{1947}).

\bibitem{kumar2022gender}
\bibinfo{author}{Kumar, A.~M.}, \bibinfo{author}{Goh, J. Y.~Q.},
  \bibinfo{author}{Tan, T. H.~H.} \& \bibinfo{author}{Siew, C. S.~Q.}
\newblock \bibinfo{title}{Gender stereotypes in hollywood movies and their
  evolution over time: Insights from network analysis}.
\newblock \emph{\bibinfo{journal}{Big Data and Cognitive Computing}}
  \textbf{\bibinfo{volume}{6}}, \bibinfo{pages}{50} (\bibinfo{year}{2022}).

\bibitem{hedley1994presentation}
\bibinfo{author}{Hedley, M.}
\newblock \bibinfo{title}{The presentation of gendered conflict in popular
  movies: Affective stereotypes, cultural sentiments, and men's motivation}.
\newblock \emph{\bibinfo{journal}{Sex Roles}} \textbf{\bibinfo{volume}{31}},
  \bibinfo{pages}{721--740} (\bibinfo{year}{1994}).

\bibitem{baker1991role}
\bibinfo{author}{Baker, W.~E.} \& \bibinfo{author}{Faulkner, R.~R.}
\newblock \bibinfo{title}{Role as resource in the hollywood film industry}.
\newblock \emph{\bibinfo{journal}{American journal of sociology}}
  \textbf{\bibinfo{volume}{97}}, \bibinfo{pages}{279--309}
  (\bibinfo{year}{1991}).

\bibitem{gerbner2002growing}
\bibinfo{author}{Gerbner, G.}, \bibinfo{author}{Gross, L.},
  \bibinfo{author}{Morgan, M.}, \bibinfo{author}{Signorielli, N.} \&
  \bibinfo{author}{Shanahan, J.}
\newblock \bibinfo{title}{Growing up with television: Cultivation processes}.
\newblock In \emph{\bibinfo{booktitle}{Media effects}}, \bibinfo{pages}{53--78}
  (\bibinfo{publisher}{Routledge}, \bibinfo{year}{2002}).

\bibitem{ward2023media}
\bibinfo{author}{Ward, L.~M.} \& \bibinfo{author}{Bridgewater, E.}
\newblock \bibinfo{title}{Media use and the development of racial attitudes
  among us youth}.
\newblock \emph{\bibinfo{journal}{Child Development Perspectives}}
  \textbf{\bibinfo{volume}{17}}, \bibinfo{pages}{83--89}
  (\bibinfo{year}{2023}).

\bibitem{tukachinsky2015documenting}
\bibinfo{author}{Tukachinsky, R.}, \bibinfo{author}{Mastro, D.} \&
  \bibinfo{author}{Yarchi, M.}
\newblock \bibinfo{title}{Documenting portrayals of race/ethnicity on primetime
  television over a 20-year span and their association with national-level
  racial/ethnic attitudes}.
\newblock \emph{\bibinfo{journal}{Journal of Social Issues}}
  \textbf{\bibinfo{volume}{71}}, \bibinfo{pages}{17--38}
  (\bibinfo{year}{2015}).

\bibitem{erigha2015race}
\bibinfo{author}{Erigha, M.}
\newblock \bibinfo{title}{Race, gender, hollywood: Representation in cultural
  production and digital media's potential for change}.
\newblock \emph{\bibinfo{journal}{Sociology compass}}
  \textbf{\bibinfo{volume}{9}}, \bibinfo{pages}{78--89} (\bibinfo{year}{2015}).

\bibitem{nwonka2020race}
\bibinfo{author}{Nwonka, C.~J.}
\newblock \bibinfo{title}{Diversity and data: an ontology of race and ethnicity
  in the british film institute’s diversity standards}.
\newblock \emph{\bibinfo{journal}{Media, Culture \& Society}}
  \textbf{\bibinfo{volume}{43}}, \bibinfo{pages}{460--479}
  (\bibinfo{year}{2021}).

\bibitem{karniouchina2023women}
\bibinfo{author}{Karniouchina, E.~V.}, \bibinfo{author}{Carson, S.~J.},
  \bibinfo{author}{Theokary, C.}, \bibinfo{author}{Rice, L.} \&
  \bibinfo{author}{Reilly, S.}
\newblock \bibinfo{title}{Women and minority film directors in hollywood:
  Performance implications of product development and distribution biases}.
\newblock \emph{\bibinfo{journal}{Journal of Marketing Research}}
  \textbf{\bibinfo{volume}{60}}, \bibinfo{pages}{25--51}
  (\bibinfo{year}{2023}).

\bibitem{borum2018oscars}
\bibinfo{author}{Borum~Chattoo, C.}
\newblock \bibinfo{title}{Oscars so white: Gender, racial, and ethnic diversity
  and social issues in us documentary films (2008--2017)}.
\newblock \emph{\bibinfo{journal}{Mass Communication and Society}}
  \textbf{\bibinfo{volume}{21}}, \bibinfo{pages}{368--394}
  (\bibinfo{year}{2018}).

\bibitem{molina2018oscarssowhite}
\bibinfo{author}{Molina-Guzm{\'a}n, I.}
\newblock \bibinfo{title}{\# oscarssowhite: how stuart hall explains why
  nothing changes in hollywood and everything is changing}.
\newblock In \emph{\bibinfo{booktitle}{Stuart Hall lives: Cultural studies in
  an age of digital media}}, \bibinfo{pages}{86--102}
  (\bibinfo{publisher}{Routledge}, \bibinfo{year}{2018}).

\bibitem{smith2017inclusion}
\bibinfo{author}{Smith, S.~L.}, \bibinfo{author}{Pieper, K.} \&
  \bibinfo{author}{Choueiti, M.}
\newblock \bibinfo{title}{Inclusion in the director’s chair? gender, race, \&
  age of film directors across 1,000 films from 2007--2016}.
\newblock \emph{\bibinfo{journal}{Media, Diversity, \& Social Change
  Initiative}}  (\bibinfo{year}{2017}).

\bibitem{ramon2023hollywood}
\bibinfo{author}{Ram{\'o}n, A.-C.}, \bibinfo{author}{Tran, M.} \&
  \bibinfo{author}{Hunt, D.}
\newblock \bibinfo{title}{Hollywood diversity report 2023}
  (\bibinfo{year}{2023}).

\bibitem{mastro2009racial}
\bibinfo{author}{Mastro, D.}
\newblock \bibinfo{title}{Racial/ethnic stereotyping and the media}.
\newblock \emph{\bibinfo{journal}{Media processes and effects}}
  \bibinfo{pages}{377--391} (\bibinfo{year}{2009}).

\bibitem{mastro2009effects}
\bibinfo{author}{Mastro, D.}
\newblock \bibinfo{title}{Effects of racial and ethnic stereotyping}.
\newblock In \emph{\bibinfo{booktitle}{Media effects}},
  \bibinfo{pages}{341--357} (\bibinfo{publisher}{Routledge},
  \bibinfo{year}{2009}).

\bibitem{mastro2005latino}
\bibinfo{author}{Mastro, D.~E.} \& \bibinfo{author}{Behm-Morawitz, E.}
\newblock \bibinfo{title}{Latino representation on primetime television}.
\newblock \emph{\bibinfo{journal}{Journalism \& Mass Communication Quarterly}}
  \textbf{\bibinfo{volume}{82}}, \bibinfo{pages}{110--130}
  (\bibinfo{year}{2005}).

\bibitem{kuppuswamy2020testing}
\bibinfo{author}{Kuppuswamy, V.} \& \bibinfo{author}{Younkin, P.}
\newblock \bibinfo{title}{Testing the theory of consumer discrimination as an
  explanation for the lack of minority hiring in hollywood films}.
\newblock \emph{\bibinfo{journal}{Management Science}}
  \textbf{\bibinfo{volume}{66}}, \bibinfo{pages}{1227--1247}
  (\bibinfo{year}{2020}).

\bibitem{topaz2022race}
\bibinfo{author}{Topaz, C.~M.} \emph{et~al.}
\newblock \bibinfo{title}{Race-and gender-based under-representation of
  creative contributors: art, fashion, film, and music}.
\newblock \emph{\bibinfo{journal}{Humanities and Social Sciences
  Communications}} \textbf{\bibinfo{volume}{9}}, \bibinfo{pages}{221}
  (\bibinfo{year}{2022}).

\bibitem{bamman2024measuring}
\bibinfo{author}{Bamman, D.}, \bibinfo{author}{Samberg, R.},
  \bibinfo{author}{So, R.~J.} \& \bibinfo{author}{Zhou, N.}
\newblock \bibinfo{title}{Measuring diversity in hollywood through the
  large-scale computational analysis of film}.
\newblock \emph{\bibinfo{journal}{Proceedings of the National Academy of
  Sciences}} \textbf{\bibinfo{volume}{121}}, \bibinfo{pages}{e2409770121}
  (\bibinfo{year}{2024}).

\bibitem{stuhler2024agency}
\bibinfo{author}{Stuhler, O.}
\newblock \bibinfo{title}{The gender agency gap in fiction writing (1850 to
  2010)}.
\newblock \emph{\bibinfo{journal}{Proceedings of the National Academy of
  Sciences}} \textbf{\bibinfo{volume}{121}}, \bibinfo{pages}{e2319514121}
  (\bibinfo{year}{2024}).

\bibitem{lauzen2020celluloid}
\bibinfo{author}{Lauzen, M.~M.}
\newblock \bibinfo{title}{The celluloid ceiling: Behind-the-scenes employment
  of women on the top 100, 250, and 500 films of 2019} (\bibinfo{year}{2020}).

\bibitem{smith2010gender}
\bibinfo{author}{Smith, S.~L.}
\newblock \bibinfo{title}{Gender oppression in cinematic content? a look at
  females on screen \& behind-the-camera in top-grossing 2007 films}.
\newblock \emph{\bibinfo{journal}{Annenberg School for Communication \&
  Journalism, University of Southern California}} \bibinfo{pages}{1--28}
  (\bibinfo{year}{2010}).

\bibitem{aldahoul2024inclusive}
\bibinfo{author}{AlDahoul, N.}, \bibinfo{author}{Ibrahim, H.},
  \bibinfo{author}{Park, M.}, \bibinfo{author}{Rahwan, T.} \&
  \bibinfo{author}{Zaki, Y.}
\newblock \bibinfo{title}{Inclusive content reduces racial and gender biases,
  yet non-inclusive content dominates popular culture}.
\newblock \emph{\bibinfo{journal}{arXiv preprint arXiv:2405.06404}}
  (\bibinfo{year}{2024}).

\bibitem{shrestha2026scene}
\bibinfo{author}{Shrestha, S.}, \bibinfo{author}{Heo, Y.},
  \bibinfo{author}{Barron, A.~T.} \& \bibinfo{author}{Park, M.}
\newblock \bibinfo{title}{Scene-level movie data from amazon x-ray in the us
  market combined with imdb}.
\newblock \emph{\bibinfo{journal}{Scientific Data}}
  \textbf{\bibinfo{volume}{13}} (\bibinfo{year}{2026}).

\bibitem{hunt2014hollywood}
\bibinfo{author}{Hunt, D.}, \bibinfo{author}{Ramon, A.-C.} \&
  \bibinfo{author}{Price, Z.}
\newblock \bibinfo{title}{Hollywood diversity report: Making sense of the
  disconnect}.
\newblock \emph{\bibinfo{journal}{Ralph Bunche Center for African American
  Studies at UCLA: http://www. bunchecenter. ucla.
  edu/wp-content/uploads/2014/02/2014-Hollywood-Diversity-Report-2-12-14. pdf}}
   (\bibinfo{year}{2014}).

\bibitem{lauzen2005maintaining}
\bibinfo{author}{Lauzen, M.~M.} \& \bibinfo{author}{Dozier, D.~M.}
\newblock \bibinfo{title}{Maintaining the double standard: Portrayals of age
  and gender in popular films}.
\newblock \emph{\bibinfo{journal}{Sex roles}} \textbf{\bibinfo{volume}{52}},
  \bibinfo{pages}{437--446} (\bibinfo{year}{2005}).

\bibitem{archer1983face}
\bibinfo{author}{Archer, D.}, \bibinfo{author}{Iritani, B.},
  \bibinfo{author}{Kimes, D.~D.} \& \bibinfo{author}{Barrios, M.}
\newblock \bibinfo{title}{Face-ism: Five studies of sex differences in facial
  prominence.}
\newblock \emph{\bibinfo{journal}{Journal of Personality and social
  Psychology}} \textbf{\bibinfo{volume}{45}}, \bibinfo{pages}{725--735}
  (\bibinfo{year}{1983}).

\bibitem{serrano2009extracting}
\bibinfo{author}{Serrano, M.~{\'A}.}, \bibinfo{author}{Bogun{\'a}, M.} \&
  \bibinfo{author}{Vespignani, A.}
\newblock \bibinfo{title}{Extracting the multiscale backbone of complex
  weighted networks}.
\newblock \emph{\bibinfo{journal}{Proceedings of the national academy of
  sciences}} \textbf{\bibinfo{volume}{106}}, \bibinfo{pages}{6483--6488}
  (\bibinfo{year}{2009}).

\bibitem{onet2025}
\bibinfo{author}{{O*NET OnLine}}.
\newblock \bibinfo{title}{{O*NET OnLine}}.
\newblock \bibinfo{howpublished}{\url{https://www.onetonline.org/}}
  (\bibinfo{year}{2025}).
\newblock \bibinfo{note}{Accessed: 2025-05-13}.

\bibitem{becker2010economics}
\bibinfo{author}{Becker, G.~S.}
\newblock \emph{\bibinfo{title}{The economics of discrimination}}
  (\bibinfo{publisher}{University of Chicago press}, \bibinfo{year}{2010}).

\bibitem{bielby1994all}
\bibinfo{author}{Bielby, W.~T.} \& \bibinfo{author}{Bielby, D.~D.}
\newblock \bibinfo{title}{" all hits are flukes": Institutionalized decision
  making and the rhetoric of network prime-time program development}.
\newblock \emph{\bibinfo{journal}{American Journal of Sociology}}
  \textbf{\bibinfo{volume}{99}}, \bibinfo{pages}{1287--1313}
  (\bibinfo{year}{1994}).

\bibitem{stroube2024status}
\bibinfo{author}{Stroube, B.~K.} \& \bibinfo{author}{Waguespack, D.~M.}
\newblock \bibinfo{title}{Status and consensus: Heterogeneity in audience
  evaluations of female-versus male-lead films}.
\newblock \emph{\bibinfo{journal}{Strategic Management Journal}}
  \textbf{\bibinfo{volume}{45}}, \bibinfo{pages}{994--1024}
  (\bibinfo{year}{2024}).

\bibitem{aguiar2026bad}
\bibinfo{author}{Aguiar, L.}
\newblock \bibinfo{title}{Bad apples on rotten tomatoes: Critics, crowds, and
  gender bias in product ratings}.
\newblock \emph{\bibinfo{journal}{Marketing Science}}
  \textbf{\bibinfo{volume}{45}}, \bibinfo{pages}{63--79}
  (\bibinfo{year}{2026}).

\bibitem{golder2022methods}
\bibinfo{author}{Golder, S.}, \bibinfo{author}{Stevens, R.},
  \bibinfo{author}{O'Connor, K.}, \bibinfo{author}{James, R.} \&
  \bibinfo{author}{Gonzalez-Hernandez, G.}
\newblock \bibinfo{title}{Methods to establish race or ethnicity of twitter
  users: scoping review}.
\newblock \emph{\bibinfo{journal}{Journal of medical Internet research}}
  \textbf{\bibinfo{volume}{24}}, \bibinfo{pages}{e35788}
  (\bibinfo{year}{2022}).

\bibitem{vanhelene2024inferring}
\bibinfo{author}{VanHelene, A.~D.} \emph{et~al.}
\newblock \bibinfo{title}{Inferring gender from first names: Comparing the
  accuracy of genderize, gender api, and the gender r package on authors of
  diverse nationality}.
\newblock \emph{\bibinfo{journal}{PLOS Digital Health}}
  \textbf{\bibinfo{volume}{3}}, \bibinfo{pages}{e0000456}
  (\bibinfo{year}{2024}).

\bibitem{lockhart2023name}
\bibinfo{author}{Lockhart, J.~W.}, \bibinfo{author}{King, M.~M.} \&
  \bibinfo{author}{Munsch, C.}
\newblock \bibinfo{title}{Name-based demographic inference and the unequal
  distribution of misrecognition}.
\newblock \emph{\bibinfo{journal}{Nature Human Behaviour}}
  \textbf{\bibinfo{volume}{7}}, \bibinfo{pages}{1084--1095}
  (\bibinfo{year}{2023}).

\bibitem{ye2017nationality}
\bibinfo{author}{Ye, J.} \emph{et~al.}
\newblock \bibinfo{title}{Nationality classification using name embeddings}.
\newblock In \emph{\bibinfo{booktitle}{Proceedings of the 2017 ACM on
  Conference on Information and Knowledge Management}},
  \bibinfo{pages}{1897--1906} (\bibinfo{year}{2017}).

\bibitem{ye2019secret}
\bibinfo{author}{Ye, J.} \& \bibinfo{author}{Skiena, S.}
\newblock \bibinfo{title}{The secret lives of names? name embeddings from
  social media}.
\newblock In \emph{\bibinfo{booktitle}{Proceedings of the 25th ACM SIGKDD
  International Conference on Knowledge Discovery \& Data Mining}},
  \bibinfo{pages}{3000--3008} (\bibinfo{year}{2019}).

\bibitem{genderize_2019}
\bibinfo{author}{Genderize.io}.
\newblock \bibinfo{title}{Genderize.io | determine the gender of a name}
  (\bibinfo{year}{2019}).
\newblock \urlprefix\url{https://genderize.io/}.

\bibitem{kempf2021partisan}
\bibinfo{author}{Kempf, E.} \& \bibinfo{author}{Tsoutsoura, M.}
\newblock \bibinfo{title}{Partisan professionals: Evidence from credit rating
  analysts}.
\newblock \emph{\bibinfo{journal}{The journal of finance}}
  \textbf{\bibinfo{volume}{76}}, \bibinfo{pages}{2805--2856}
  (\bibinfo{year}{2021}).

\bibitem{luca2024evolution}
\bibinfo{author}{Luca, M.}, \bibinfo{author}{Pronkina, E.} \&
  \bibinfo{author}{Rossi, M.}
\newblock \bibinfo{title}{The evolution of discrimination in online markets:
  How the rise in anti-asian bias affected airbnb during the pandemic}.
\newblock \emph{\bibinfo{journal}{Marketing Science}}
  \textbf{\bibinfo{volume}{45}}, \bibinfo{pages}{108--122}
  (\bibinfo{year}{2026}).

\bibitem{openai2024gpt4v}
\bibinfo{author}{{OpenAI}}.
\newblock \bibinfo{title}{{GPT-4} with vision} (\bibinfo{year}{2024}).
\newblock \urlprefix\url{https://platform.openai.com}.

\bibitem{fasttext}
\bibinfo{author}{{FastText}}.
\newblock \bibinfo{title}{English word vectors}.
\newblock \urlprefix\url{https://fasttext.cc/docs/en/english-vectors.html}.

\end{thebibliography}

\newpage

\section*{Acknowledgements}
M.P. was partially supported by the Center for Interdisciplinary Data Science and Artificial Intelligence(CIDSAI), funded by Tamkeen under NYUAD Research Institute awards CG016. The authors disclose the use of generative AI in the research and writing process. Following the GAIDeT taxonomy (2025), code generation and proofreading and editing were delegated to generative AI tools (Claude Opus/Sonnet 4.6) on a limited basis under full human supervision. For example, initial figure-generation scripts were written by hand, with generative AI subsequently used only to organize the code into a public-facing GitHub repository. Responsibility for the final manuscript lies entirely with the authors, and generative AI tools are not listed as authors and bear no responsibility for the final outcomes.

\section*{Data and code availability}
All code needed to reproduce the figures and statistical analyses in this paper is publicly available at \url{https://github.com/comnetsAD/hollywood_onscreen_inertia}, comprising self-contained Python scripts (one per figure) together with the aggregated, role-level data tables they consume. The smaller derived tables used by the analyses---decennial U.S.\ Census population shares, the O*NET occupational-field mapping, film metadata, budget and revenue records, and scene-level co-appearance data---are included directly in the repository; the larger role-level tables that exceed GitHub's file-size limit are hosted in a public Google Drive folder linked from the repository (\url{https://drive.google.com/drive/folders/1riImQuBLUkMPtoH3IhukU5i7CloIBgbh}).

The underlying records were assembled from third-party sources, each available under its own terms of use. Film metadata, cast and crew credits, posters, budgets, revenues, and audience ratings were retrieved from The Movie Database (TMDB) API (\url{https://www.themoviedb.org}); this product uses the TMDB API but is not endorsed or certified by TMDB. Scene-level character co-appearance data are from the augmented Amazon X-Ray dataset of Shrestha et al.~\cite{shrestha2026scene}. Perceived race and gender were inferred from names using NamePrism~\cite{ye2017nationality, ye2019secret} and Genderize~\cite{genderize_2019}, respectively; occupational roles were mapped to the O*NET taxonomy~\cite{onet2025}; and movie-poster portrayals were classified using the GPT-4V pipeline of AlDahoul et al.~\cite{aldahoul2024inclusive}.

\section*{Author Contributions}
M.P. conceived the study; M.P. and H.I. designed the research with input from Y.Z. and T.R.; H.I. curated and collected data, performed research, and prepared visualizations; M.P. and H.I. analyzed data and drafted and revised the manuscript; all authors discussed the results and edited the paper; and Y.Z. and M.P. supervised the study with guidance from T.R.

\section*{Competing Interests}
The authors declare no competing interests.

\clearpage

\setcounter{figure}{0}
\setcounter{table}{0}
\renewcommand{\figurename}{Supplementary Figure}
\renewcommand{\tablename}{Supplementary Table}

\begin{center}
{\fontsize{14}{14}\selectfont Supplementary Information for}

\medskip

{\fontsize{16}{16}\selectfont On-Screen Inertia: Persistent Racial and Gender Disparities in Hollywood Film (1900-2024)}

\medskip

Hazem Ibrahim$^{1}$, Talal Rahwan$^{1}$, Yasir Zaki$^{1,*}$, Minsu Park$^{1,*}$

\medskip

$^{1}$New York University Abu Dhabi, UAE

$^{*}$To whom correspondence should be addressed; E-mail: \{yasir.zaki, minsu.park\}@nyu.edu
\end{center}

\bigskip

This document is structured as follows:
\smallskip\smallskip
\begingroup
\renewcommand{\labelitemi}{--}
\begin{itemize}\itemsep0.5em
\item \textbf{Supplementary Note 1:} Age analysis of racial and gender groups  (\emph{page~\pageref{age}})
\item \textbf{Supplementary Tables~1 to 17} (\emph{page~\pageref{tables}})
\item \textbf{Supplementary Figures~1 to 15} (\emph{page~\pageref{figures}})
\end{itemize}
\endgroup

\clearpage

\section*{Supplementary Note 1: Age analysis of racial and gender groups}
\label{age}

\begin{figure}[htbp!]
    \centering
    \includegraphics[width=\linewidth]{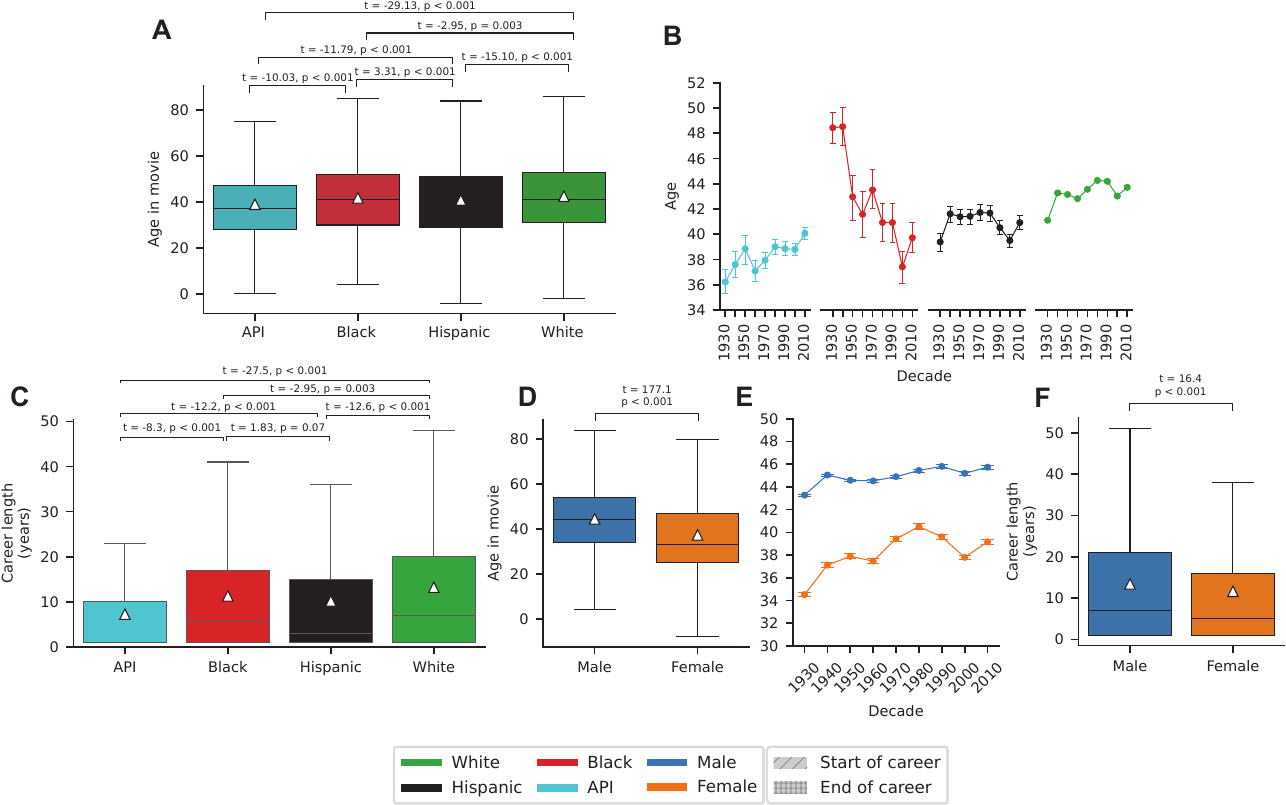}
        \caption{
        \textbf{Analysis of cast member ages across racial and gender groups.} (\textbf{A}) Distribution of appearance ages for each racial group; triangles denote group means. (\textbf{B}) Mean appearance age by decade for each racial group. (\textbf{C}) Distribution of career lengths for each racial group. (\textbf{D}--\textbf{F}) Equivalent analyses for gender groups. All pairwise comparisons use independent two-sample $t$-tests; test statistics are annotated on each panel.
    }
    \label{fig:age_analysis}
\end{figure}

The main text reports age and career-length disparities between male and female actors
and notes parallel patterns across racial groups (Figure~1D,E). This note extends
those analyses across all demographic dimensions. Supplementary Figure~\ref{fig:age_analysis}
presents three matched comparisons, each examined separately for racial groups
(Panels A--C) and gender groups (Panels D--F): the distribution of actor ages at the
time of on-screen appearance (Panels A and D), how mean appearance age has shifted
across decades (Panels B and E), and the distribution of career lengths (Panels C and F).
Note that birth year data are not available for all actors; see Supplementary
Table~\ref{tab:birthyear_na_decade} for missingness rates by decade.

Across racial groups, we observe clear age stratification at the moment of on-screen appearance (Supplementary Figure~\ref{fig:age_analysis}A). White actors are, on average, the oldest when they appear in films (mean age $= 42.56$), followed by Black actors ($41.78$), Hispanic actors ($40.86$), and Asian/Pacific Islander (API) actors ($39.00$). All pairwise comparisons are statistically significant. These differences are not static; they vary across decades (Supplementary Figure~\ref{fig:age_analysis}B). For API and White actors, average age increases slightly between the 1930s and 2010s (from approximately $36$ to $40$ for API actors, and from approximately $41$ to $43$ for White actors). Hispanic actors remain comparatively stable over time. In contrast, Black actors exhibit a sharp decline, from an average age of approximately $49$ in the 1930s--1940s to approximately $40$ in the 2010s.

To better interpret these aggregate age patterns, we next examine age differences across the arc of an actor's observed career in the dataset. Main Figure~1D contrasts the age at which an actor first appears (career entry) with the age at which they last appear (career exit). We find that White actors tend to enter Hollywood later and exit at older ages (career start: 34.45 years; career end: 45.95 years). Hispanic actors exhibit a similar age-at-entry profile (34.51 years) but exit earlier on average (43.41 years). In contrast, Black actors enter at younger ages (32.87 years) yet exit at an average age of 42.71 years, while API actors have the youngest career-end ages overall (career start: 33.63 years; career end: 39.74 years). The gap between entry and exit ages shows corresponding differences in observed career length (Supplementary Figure~\ref{fig:age_analysis}C). White actors have the longest careers on average (13.27 years), followed by Black actors (11.32 years) and Hispanic actors (10.24 years), with API actors exhibiting the shortest observed careers (7.31 years).

Gender groups show a parallel structure of inequalities, though the patterns differ in both magnitude and direction. Male actors are older on average than female actors at the time of appearance (Supplementary Figure~\ref{fig:age_analysis}D; $\Delta = 7.13$, $t = 177.1$, $p < 0.001$), indicating that women are generally cast at younger ages. Both male and female actors' mean appearance ages have increased slightly across the span of the dataset (Supplementary Figure~\ref{fig:age_analysis}E), suggesting a broad industry-wide shift toward older casts in the 21st century.

Finally, career trajectories indicate that these gender gaps persist at both entry and exit (Main Figure~1E): female actors enter and leave at younger ages than male actors (Career start: $\Delta = 5.48$, $t = 54.1$, $p < 0.001$; Career end: $\Delta = 6.97$, $t = 55.1$, $p < 0.001$). However, the gap between entry and exit ages is smaller for women, which manifests as shorter observed careers, as seen in Supplementary Figure~\ref{fig:age_analysis}F (Career length: $\Delta = 1.77$, $t = 16.4$, $p < 0.001$). This pattern is consistent with the well-documented ``double standard'' of aging in entertainment, wherein longevity and continued visibility are more strongly afforded to men than to women---a further dimension of on-screen inertia, as the industry's tendency to cast women at younger ages and afford them shorter careers has persisted across decades, consistent with and reflective of broader cultural norms around gender and aging.

\clearpage

\section*{Supplementary Tables}
\label{tables}

\begin{table}[htbp]
\centering
\small
\caption{Number of films and credited cast and crew by decade.}
\label{tab:film_counts_per_decade}
\begin{tabular}{lrrr}
\toprule
Decade &
\begin{tabular}[c]{@{}r@{}}Number of\\films\end{tabular} &
\begin{tabular}[c]{@{}r@{}}Number of\\cast\\credits\end{tabular} &
\begin{tabular}[c]{@{}r@{}}Number of\\crew\\credits\end{tabular} \\
\midrule
1900s & 2112 & 329 & 169 \\
1910s & 8399 & 5729 & 1778 \\
1920s & 8550 & 6294 & 3053 \\
1930s & 11282 & 10754 & 5536 \\
1940s & 10211 & 10303 & 5828 \\
1950s & 9050 & 11939 & 7621 \\
1960s & 11573 & 18615 & 12039 \\
1970s & 17068 & 32754 & 22127 \\
1980s & 20394 & 54581 & 39841 \\
1990s & 30187 & 74400 & 64401 \\
2000s & 59486 & 129550 & 116730 \\
2010s & 96881 & 245386 & 245199 \\
2020s & 47200 & 154444 & 175487 \\
\bottomrule
\end{tabular}
\end{table}

\begin{table}[htbp]
\centering
\small
\caption{Share of characters whose gender and race could be classified at different confidence thresholds. Shares are measured between 0 and 1.}
\label{tab:gender_race_conf_thresholds}
\begin{tabular}{lrr}
\toprule
\begin{tabular}[c]{@{}l@{}}Confidence\\threshold\end{tabular} &
\begin{tabular}[c]{@{}r@{}}Share of characters\\passing gender\\confidence\end{tabular} &
\begin{tabular}[c]{@{}r@{}}Share of characters\\passing race\\confidence\end{tabular} \\
\midrule
0.1 & 0.987529 & 1.000000 \\
0.2 & 0.987529 & 1.000000 \\
0.3 & 0.987529 & 0.999995 \\
0.4 & 0.987529 & 0.999234 \\
0.5 & 0.987529 & 0.991875 \\
0.6 & 0.974438 & 0.967247 \\
0.7 & 0.955620 & 0.938088 \\
0.8 & 0.928955 & 0.890136 \\
0.9 & 0.891926 & 0.750693 \\
1.0 & 0.668182 & 0.000000 \\
\bottomrule
\end{tabular}
\end{table}

\begin{table}[htbp]
\centering
\small
\caption{Missing birth year information for roles and actors, by decade. Shares are measured between 0 and 1.}
\label{tab:birthyear_na_decade}
\begin{tabular}{lrrrrrrr}
\toprule
Decade &
\begin{tabular}[c]{@{}r@{}}Total\\roles\end{tabular} &
\begin{tabular}[c]{@{}r@{}}Unique\\actors\end{tabular} &
\begin{tabular}[c]{@{}r@{}}Roles with\\missing\\birth year\end{tabular} &
\begin{tabular}[c]{@{}r@{}}Actors with\\missing\\birth year\end{tabular} &
\begin{tabular}[c]{@{}r@{}}Share of roles\\with missing\\birth year\end{tabular} &
\begin{tabular}[c]{@{}r@{}}Share of actors\\with missing\\birth year\end{tabular} \\
\midrule
1900s & 2253 & 524 & 332 & 279 & 0.147359 & 0.532443 \\
1910s & 42163 & 8094 & 7600 & 4432 & 0.180253 & 0.547566 \\
1920s & 50226 & 10209 & 7794 & 5120 & 0.155179 & 0.501518 \\
1930s & 121306 & 16675 & 15624 & 8785 & 0.128798 & 0.526837 \\
1940s & 114419 & 17583 & 16567 & 8749 & 0.144792 & 0.497583 \\
1950s & 88124 & 21348 & 15347 & 10005 & 0.174152 & 0.468662 \\
1960s & 78976 & 31747 & 22047 & 16269 & 0.279161 & 0.512458 \\
1970s & 121492 & 53141 & 42283 & 31868 & 0.348031 & 0.599688 \\
1980s & 181057 & 85203 & 77700 & 57332 & 0.429147 & 0.672887 \\
1990s & 256749 & 116145 & 112662 & 80058 & 0.438802 & 0.689294 \\
2000s & 380655 & 188893 & 183070 & 136439 & 0.480934 & 0.722308 \\
2010s & 610428 & 317063 & 359301 & 253314 & 0.588605 & 0.798939 \\
2020s & 318800 & 206933 & 214960 & 165974 & 0.674279 & 0.802066 \\
\bottomrule
\end{tabular}
\end{table}

\begin{sidewaystable}[htbp]
\centering
\scriptsize
\caption{Number of roles by occupational field and decade.}
\label{tab:field_counts}
\begin{tabular}{lrrrrrrrrrrrrr}
\toprule
\begin{tabular}[c]{@{}l@{}}Occupational\\field\end{tabular} &
1900s & 1910s & 1920s & 1930s & 1940s & 1950s & 1960s &
1970s & 1980s & 1990s & 2000s & 2010s & 2020s \\
\midrule
Arts, Design, Entertainment, Sports, and Media & 1 & 20 & 25 & 569 & 548 & 160 & 217 & 604 & 1120 & 1263 & 1663 & 2737 & 817 \\
Building, Grounds Cleaning, and Maintenance & 0 & 23 & 23 & 67 & 47 & 11 & 25 & 52 & 79 & 81 & 113 & 159 & 60 \\
Business and Financial Operations & 0 & 11 & 14 & 84 & 136 & 47 & 64 & 143 & 195 & 222 & 272 & 508 & 182 \\
Community and Social Services & 0 & 1 & 1 & 1 & 0 & 0 & 2 & 10 & 27 & 49 & 54 & 133 & 66 \\
Computer and Mathematics & 0 & 0 & 0 & 1 & 2 & 3 & 5 & 7 & 23 & 45 & 55 & 115 & 44 \\
Construction and Extraction & 0 & 1 & 1 & 20 & 13 & 8 & 13 & 21 & 43 & 49 & 87 & 247 & 105 \\
Education, Training, and Library Services & 1 & 2 & 1 & 16 & 28 & 4 & 17 & 50 & 118 & 121 & 198 & 400 & 122 \\
Engineering and Architecture & 0 & 0 & 0 & 1 & 0 & 1 & 0 & 2 & 4 & 2 & 7 & 14 & 4 \\
Farming, Fishing, and Forestry & 1 & 3 & 5 & 70 & 111 & 17 & 19 & 78 & 81 & 63 & 83 & 264 & 83 \\
Food Preparation and Serving & 0 & 12 & 13 & 116 & 164 & 36 & 70 & 136 & 213 & 323 & 352 & 793 & 256 \\
Healthcare & 0 & 9 & 19 & 153 & 146 & 79 & 106 & 311 & 489 & 637 & 646 & 1218 & 313 \\
Installation, Maintenance, and Repair & 0 & 1 & 1 & 24 & 21 & 5 & 5 & 28 & 43 & 47 & 45 & 92 & 32 \\
Law and Legal Services & 0 & 15 & 8 & 62 & 74 & 14 & 23 & 57 & 107 & 122 & 126 & 267 & 67 \\
Life, Physical, and Social Science & 0 & 0 & 2 & 5 & 5 & 5 & 6 & 14 & 17 & 24 & 29 & 72 & 33 \\
Management & 0 & 12 & 11 & 33 & 72 & 30 & 19 & 88 & 71 & 97 & 91 & 215 & 63 \\
Manufacturing and Production & 0 & 2 & 4 & 38 & 34 & 13 & 42 & 76 & 91 & 225 & 151 & 300 & 104 \\
Military & 0 & 0 & 0 & 9 & 0 & 0 & 0 & 1 & 3 & 1 & 1 & 12 & 8 \\
Office and Administrative Support & 0 & 31 & 34 & 154 & 163 & 51 & 109 & 234 & 368 & 399 & 483 & 1009 & 268 \\
Personal Care and Related Services & 0 & 5 & 6 & 48 & 66 & 23 & 29 & 59 & 89 & 117 & 78 & 178 & 50 \\
Protective Services & 0 & 61 & 65 & 406 & 384 & 104 & 121 & 290 & 422 & 568 & 376 & 776 & 217 \\
Sales and Related Services & 0 & 3 & 0 & 8 & 11 & 4 & 4 & 22 & 55 & 68 & 67 & 190 & 74 \\
Transportation and Material Moving & 1 & 18 & 18 & 137 & 150 & 46 & 76 & 140 & 159 & 212 & 202 & 387 & 133 \\
\bottomrule
\end{tabular}
\end{sidewaystable}

\begin{table}[htbp]
\centering
\small
\caption{Normalized betweenness centrality differences across demographic groups for varying values of $\alpha$.}
\label{tab:alpha_sweep_betweenness}
\begin{tabular}{lrrrrrr}
\toprule
$\alpha$ & API & Black & Hispanic & White & Female & Male \\
\midrule
0.05 & -0.0172 & -0.0120 & -0.0421 &  0.0025 & -0.0081 &  0.0047 \\
0.10 & -0.0319 & -0.0278 & -0.0430 &  0.0033 & -0.0034 &  0.0018 \\
0.15 & -0.0224 & -0.0200 & -0.0350 &  0.0027 &  0.0005 & -0.0003 \\
0.20 & -0.0268 & -0.0172 & -0.0243 &  0.0022 &  0.0020 & -0.0011 \\
0.25 & -0.0248 & -0.0064 & -0.0210 &  0.0019 &  0.0012 & -0.0006 \\
0.50 & -0.0197 & -0.0058 & -0.0154 &  0.0014 &  0.0015 & -0.0008 \\
1.00 & -0.0173 & -0.0047 & -0.0134 &  0.0012 &  0.0014 & -0.0007 \\
\bottomrule
\end{tabular}
\end{table}

\begin{table}[htbp]
\centering
\small
\caption{Normalized closeness centrality differences across demographic groups for varying values of $\alpha$.}
\label{tab:alpha_sweep_closeness}
\begin{tabular}{lrrrrrr}
\toprule
$\alpha$ & API & Black & Hispanic & White & Female & Male \\
\midrule
0.05 & -0.0026 & -0.0698 & -0.0191 &  0.0021 & -0.0015 &  0.0009 \\
0.10 & -0.0210 & -0.0251 & -0.0325 &  0.0025 &  0.0043 & -0.0024 \\
0.15 & -0.0198 & -0.0210 & -0.0377 &  0.0027 &  0.0054 & -0.0030 \\
0.20 & -0.0118 & -0.0184 & -0.0218 &  0.0017 &  0.0059 & -0.0031 \\
0.25 & -0.0152 & -0.0139 & -0.0157 &  0.0015 &  0.0046 & -0.0024 \\
0.50 & -0.0127 & -0.0111 & -0.0132 &  0.0013 &  0.0042 & -0.0022 \\
1.00 & -0.0109 & -0.0098 & -0.0115 &  0.0012 &  0.0038 & -0.0020 \\
\bottomrule
\end{tabular}
\end{table}

\begin{table}[htbp]
\centering
\small
\caption{Normalized degree centrality differences across demographic groups for varying values of $\alpha$.}
\label{tab:alpha_sweep_degree}
\begin{tabular}{lrrrrrr}
\toprule
$\alpha$ & API & Black & Hispanic & White & Female & Male \\
\midrule
0.05 & -0.0112 & -0.0962 & -0.0275 &  0.0031 & -0.0080 &  0.0045 \\
0.10 & -0.0313 & -0.0405 & -0.0411 &  0.0034 & -0.0005 &  0.0002 \\
0.15 & -0.0361 & -0.0214 & -0.0448 &  0.0035 &  0.0029 & -0.0016 \\
0.20 & -0.0363 & -0.0300 & -0.0335 &  0.0032 &  0.0053 & -0.0027 \\
0.25 & -0.0360 & -0.0181 & -0.0320 &  0.0030 &  0.0043 & -0.0022 \\
0.50 & -0.0334 & -0.0156 & -0.0287 &  0.0027 &  0.0039 & -0.0020 \\
1.00 & -0.0318 & -0.0139 & -0.0261 &  0.0025 &  0.0036 & -0.0018 \\
\bottomrule
\end{tabular}
\end{table}

\begin{table}[htbp]
\centering
\small
\caption{Star ratio by gender pairing for varying values of $\alpha$. Columns correspond to Star-Actor pair (e.g. Female star - Male non-star = F-M)}
\label{tab:star_ratio_alpha_sweep}
\begin{tabular}{lrrrr}
\toprule
$\alpha$ & F--F & F--M & M--F & M--M \\
\midrule
0.05 & 0.4307 & 0.4289 & 0.4591 & 0.3830 \\
0.10 & 0.3645 & 0.3475 & 0.3945 & 0.3319 \\
0.15 & 0.3218 & 0.3033 & 0.3211 & 0.2941 \\
0.20 & 0.2955 & 0.2731 & 0.2751 & 0.2633 \\
0.25 & 0.2615 & 0.2461 & 0.2469 & 0.2371 \\
0.50 & 0.2408 & 0.2265 & 0.2271 & 0.2189 \\
1.00 & 0.2236 & 0.2104 & 0.2109 & 0.2031 \\
\bottomrule
\end{tabular}
\end{table}

\begin{table}[htbp]
\centering
\small
\caption{Director--actor race normalized pairing likelihood for different Top-$N$ cast order cutoffs.}
\label{tab:director_actor_race_pairing}
\begin{tabular}{llrrrrr}
\toprule
\begin{tabular}{@{}l@{}}Director\\race\end{tabular} &
\begin{tabular}{@{}l@{}}Actor\\race\end{tabular} &
\begin{tabular}{@{}c@{}}Top\\3\end{tabular} &
\begin{tabular}{@{}c@{}}Top\\5\end{tabular} &
\begin{tabular}{@{}c@{}}Top\\10\end{tabular} &
\begin{tabular}{@{}c@{}}Top\\20\end{tabular} &
\begin{tabular}{@{}c@{}}Top\\30\end{tabular} \\
\midrule
API       & API       &  0.1085 &  0.1580 &  0.2103 &  0.2224 &  0.2254 \\
API       & Black     & -0.2586 & -0.2693 & -0.2854 & -0.3010 & -0.3019 \\
API       & Hispanic  & -0.1987 & -0.1765 & -0.1377 & -0.1225 & -0.1223 \\
API       & White     &  0.5763 &  0.6138 &  0.6054 &  0.5733 &  0.5607 \\
\midrule
Black     & API       & -0.2497 & -0.2267 & -0.2027 & -0.1774 & -0.1774 \\
Black     & Black     & -0.0730 & -0.0461 & -0.0380 & -0.0210 & -0.0246 \\
Black     & Hispanic  & -0.2275 & -0.2130 & -0.1890 & -0.1660 & -0.1711 \\
Black     & White     &  0.6877 &  0.6851 &  0.6516 &  0.6152 &  0.6028 \\
\midrule
Hispanic  & API       & -0.2175 & -0.1965 & -0.1656 & -0.1473 & -0.1456 \\
Hispanic  & Black     & -0.2562 & -0.2659 & -0.2886 & -0.3057 & -0.3121 \\
Hispanic  & Hispanic  &  0.0961 &  0.1715 &  0.2439 &  0.2626 &  0.2639 \\
Hispanic  & White     &  0.6078 &  0.6361 &  0.6251 &  0.5934 &  0.5817 \\
\midrule
White     & API       & -0.2471 & -0.2434 & -0.2302 & -0.2122 & -0.2059 \\
White     & Black     & -0.2622 & -0.2700 & -0.2805 & -0.2882 & -0.2892 \\
White     & Hispanic  & -0.2182 & -0.2013 & -0.1693 & -0.1426 & -0.1344 \\
White     & White     &  0.7126 &  0.6928 &  0.6537 &  0.6176 &  0.6051 \\
\bottomrule
\end{tabular}
\end{table}

\begin{table}[htbp]
\centering
\small
\caption{Director--actor gender normalized pairing likelihood for different Top-$N$ cast order cutoffs.}
\label{tab:director_actor_gender_pairing}
\begin{tabular}{llrrrrr}
\toprule
\begin{tabular}{@{}l@{}}Director\\gender\end{tabular} &
\begin{tabular}{@{}l@{}}Actor\\gender\end{tabular} &
\begin{tabular}{@{}c@{}}Top\\3\end{tabular} &
\begin{tabular}{@{}c@{}}Top\\5\end{tabular} &
\begin{tabular}{@{}c@{}}Top\\10\end{tabular} &
\begin{tabular}{@{}c@{}}Top\\20\end{tabular} &
\begin{tabular}{@{}c@{}}Top\\30\end{tabular} \\
\midrule
Female & Female & -0.0168 & -0.0094 & -0.0051 & -0.0044 & -0.0044 \\
Female & Male   &  0.0023 &  0.0058 &  0.0050 &  0.0023 &  0.0022 \\
\midrule
Male   & Female & -0.0827 & -0.0414 & -0.0178 & -0.0126 & -0.0122 \\
Male   & Male   &  0.0847 &  0.0419 &  0.0178 &  0.0129 &  0.0125 \\
\bottomrule
\end{tabular}
\end{table}

\begin{table}[htbp]
\centering
\small
\caption{Differences in normalized degree centrality for director--actor gender majority pairing across Top-$N$ characters.}
\label{tab:gg_deg_top_N}
\begin{tabular}{llrrrrrr}
\toprule
\begin{tabular}{@{}c@{}}Director\\majority\end{tabular} &
\begin{tabular}{@{}c@{}}Actor\\majority\end{tabular} &
\begin{tabular}{@{}c@{}}Top 3\end{tabular} &
\begin{tabular}{@{}c@{}}Top 5\end{tabular} &
\begin{tabular}{@{}c@{}}Top 10\end{tabular} &
\begin{tabular}{@{}c@{}}Top 20\end{tabular} &
\begin{tabular}{@{}c@{}}Top 30\end{tabular} &
\begin{tabular}{@{}c@{}}Top 50\end{tabular} \\
\midrule
Female & Female &  0.0659 &  0.0430 &  0.0265 &  0.0256 &  0.0260 &  0.0246 \\
Female & Male   & -0.0562 & -0.0349 & -0.0221 & -0.0160 & -0.0152 & -0.0144 \\
Male   & Female & -0.0045 & -0.0017 &  0.0010 &  0.0046 &  0.0055 &  0.0057 \\
Male   & Male   &  0.0008 &  0.0028 &  0.0027 & -0.0006 & -0.0014 & -0.0015 \\
\bottomrule
\end{tabular}
\end{table}

\begin{table}[htbp]
\centering
\small
\caption{Differences in normalized closeness centrality for director--actor gender majority pairing across Top-$N$ characters.}
\label{tab:gg_cls_top_N}
\begin{tabular}{llrrrrrr}
\toprule
\begin{tabular}{@{}c@{}}Director\\majority\end{tabular} &
\begin{tabular}{@{}c@{}}Actor\\majority\end{tabular} &
\begin{tabular}{@{}c@{}}Top 3\end{tabular} &
\begin{tabular}{@{}c@{}}Top 5\end{tabular} &
\begin{tabular}{@{}c@{}}Top 10\end{tabular} &
\begin{tabular}{@{}c@{}}Top 20\end{tabular} &
\begin{tabular}{@{}c@{}}Top 30\end{tabular} &
\begin{tabular}{@{}c@{}}Top 50\end{tabular} \\
\midrule
Female & Female &  0.0326 &  0.0231 &  0.0178 &  0.0165 &  0.0154 &  0.0161 \\
Female & Male   & -0.0298 & -0.0219 & -0.0130 & -0.0108 & -0.0096 & -0.0101 \\
Male   & Female & -0.0028 & -0.0012 &  0.0015 &  0.0032 &  0.0030 &  0.0033 \\
Male   & Male   &  0.0002 &  0.0011 &  0.0013 & -0.0008 & -0.0009 & -0.0010 \\
\bottomrule
\end{tabular}
\end{table}

\begin{table}[htbp]
\centering
\small
\caption{Differences in normalized betweenness centrality for director--actor gender majority pairing across Top-$N$ characters.}
\label{tab:gg_bet_top_N}
\begin{tabular}{llrrrrrr}
\toprule
\begin{tabular}{@{}c@{}}Director\\majority\end{tabular} &
\begin{tabular}{@{}c@{}}Actor\\majority\end{tabular} &
\begin{tabular}{@{}c@{}}Top 3\end{tabular} &
\begin{tabular}{@{}c@{}}Top 5\end{tabular} &
\begin{tabular}{@{}c@{}}Top 10\end{tabular} &
\begin{tabular}{@{}c@{}}Top 20\end{tabular} &
\begin{tabular}{@{}c@{}}Top 30\end{tabular} &
\begin{tabular}{@{}c@{}}Top 50\end{tabular} \\
\midrule
Female & Female &  0.0267 &  0.0195 &  0.0127 &  0.0105 &  0.0104 &  0.0105 \\
Female & Male   & -0.0231 & -0.0144 & -0.0059 & -0.0038 & -0.0034 & -0.0036 \\
Male   & Female & -0.0036 & -0.0026 & -0.0008 &  0.0000 &  0.0007 &  0.0006 \\
Male   & Male   &  0.0020 &  0.0021 &  0.0014 &  0.0002 & -0.0003 & -0.0001 \\
\bottomrule
\end{tabular}
\end{table}

\begin{table}[htbp]
\centering
\small
\caption{Differences in normalized degree centrality for director--actor race majority pairing across Top-$N$ characters.}
\label{tab:rr_deg_top_N}
\begin{tabular}{llrrrrrr}
\toprule
\begin{tabular}{@{}c@{}}Director\\majority\end{tabular} &
\begin{tabular}{@{}c@{}}Actor\\majority\end{tabular} &
\begin{tabular}{@{}c@{}}Top 3\end{tabular} &
\begin{tabular}{@{}c@{}}Top 5\end{tabular} &
\begin{tabular}{@{}c@{}}Top 10\end{tabular} &
\begin{tabular}{@{}c@{}}Top 20\end{tabular} &
\begin{tabular}{@{}c@{}}Top 30\end{tabular} &
\begin{tabular}{@{}c@{}}Top 50\end{tabular} \\
\midrule
Non-White & Non-White & -0.0071 & -0.0140 &  0.0012 & -0.0014 & -0.0033 & -0.0089 \\
Non-White & White     &  0.0083 &  0.0043 & -0.0006 &  0.0013 &  0.0016 &  0.0023 \\
White     & Non-White & -0.0357 & -0.0333 & -0.0245 & -0.0208 & -0.0232 & -0.0231 \\
White     & White     &  0.0037 &  0.0038 &  0.0027 &  0.0021 &  0.0022 &  0.0021 \\
\bottomrule
\end{tabular}
\end{table}

\begin{table}[htbp]
\centering
\small
\caption{Differences in normalized closeness centrality for director--actor race majority pairing across Top-$N$ characters.}
\label{tab:rr_cls_top_N}
\begin{tabular}{llrrrrrr}
\toprule
\begin{tabular}{@{}c@{}}Director\\majority\end{tabular} &
\begin{tabular}{@{}c@{}}Actor\\majority\end{tabular} &
\begin{tabular}{@{}c@{}}Top 3\end{tabular} &
\begin{tabular}{@{}c@{}}Top 5\end{tabular} &
\begin{tabular}{@{}c@{}}Top 10\end{tabular} &
\begin{tabular}{@{}c@{}}Top 20\end{tabular} &
\begin{tabular}{@{}c@{}}Top 30\end{tabular} &
\begin{tabular}{@{}c@{}}Top 50\end{tabular} \\
\midrule
Non-White & Non-White & -0.0086 & -0.0053 & -0.0025 & -0.0081 & -0.0043 & -0.0044 \\
Non-White & White     &  0.0075 &  0.0029 &  0.0010 &  0.0026 &  0.0026 &  0.0026 \\
White     & Non-White & -0.0228 & -0.0194 & -0.0135 & -0.0106 & -0.0131 & -0.0129 \\
White     & White     &  0.0024 &  0.0023 &  0.0016 &  0.0012 &  0.0013 &  0.0013 \\
\bottomrule
\end{tabular}
\end{table}

\begin{table}[htbp]
\centering
\small
\caption{Differences in normalized betweenness centrality for director--actor race majority pairing across Top-$N$ characters.}
\label{tab:rr_bet_top_N}
\begin{tabular}{llrrrrrr}
\toprule
\begin{tabular}{@{}c@{}}Director\\majority\end{tabular} &
\begin{tabular}{@{}c@{}}Actor\\majority\end{tabular} &
\begin{tabular}{@{}c@{}}Top 3\end{tabular} &
\begin{tabular}{@{}c@{}}Top 5\end{tabular} &
\begin{tabular}{@{}c@{}}Top 10\end{tabular} &
\begin{tabular}{@{}c@{}}Top 20\end{tabular} &
\begin{tabular}{@{}c@{}}Top 30\end{tabular} &
\begin{tabular}{@{}c@{}}Top 50\end{tabular} \\
\midrule
Non-White & Non-White &  0.0044 &  0.0025 &  0.0034 & -0.0012 & -0.0012 & -0.0008 \\
Non-White & White     &  0.0006 & -0.0016 & -0.0006 &  0.0002 &  0.0002 &  0.0002 \\
White     & Non-White & -0.0181 & -0.0137 & -0.0088 & -0.0069 & -0.0076 & -0.0076 \\
White     & White     &  0.0020 &  0.0017 &  0.0012 &  0.0009 &  0.0009 &  0.0009 \\
\bottomrule
\end{tabular}
\end{table}

\begin{table}[htbp]
\centering
\caption{Pooled OLS regression results: Cast and crew diversity on profit and audience ratings. All models use HC3 robust standard errors and control for log(budget), log(runtime), log(cast size), log(crew size), genre fixed effects, and a cross-diversity control. M1: decade fixed effects; M2: year as a continuous control; M3: decade fixed effects with crew \% White and crew \% Female as additional controls (cast variables only); M4: year as a continuous control with crew diversity controls (cast variables only). Significance: $^{*}p<0.05$, $^{**}p<0.01$, $^{***}p<0.001$.}
\label{tab:pooled_regression}
\small
\begin{tabular}{llrrrl}
\hline
\textbf{Model} & \textbf{Variable} & \textbf{$\beta$} & \textbf{95\% CI} & \textbf{$p$} & \textbf{$n$} \\
\hline
\multicolumn{6}{l}{\textit{Profit (\$M)}} \\
\hline
M1 & Cast \% White & $-$5.41 & [$-$22.19, 11.38] & 0.528 & 8,298 \\
M1 & Crew \% White & $-$30.24$^{**}$ & [$-$52.14, $-$8.34] & 0.007 & 8,129 \\
M1 & Cast \% Female & 9.38 & [$-$7.93, 26.70] & 0.288 & 8,298 \\
M1 & Crew \% Female & $-$45.80$^{***}$ & [$-$67.76, $-$23.84] & $<$0.001 & 8,129 \\
M2 & Cast \% White & $-$2.85 & [$-$19.67, 13.97] & 0.740 & 8,298 \\
M2 & Crew \% White & $-$28.71$^{**}$ & [$-$50.24, $-$7.19] & 0.009 & 8,129 \\
M2 & Cast \% Female & 8.37 & [$-$8.96, 25.70] & 0.344 & 8,298 \\
M2 & Crew \% Female & $-$57.57$^{***}$ & [$-$79.48, $-$35.67] & $<$0.001 & 8,129 \\
M3 & Cast \% White & 11.20 & [$-$12.41, 34.81] & 0.353 & 8,058 \\
M3 & Cast \% Female & 27.06$^{**}$ & [7.45, 46.66] & 0.007 & 8,058 \\
M4 & Cast \% White & 15.02 & [$-$8.66, 38.70] & 0.214 & 8,058 \\
M4 & Cast \% Female & 28.38$^{**}$ & [8.81, 47.96] & 0.004 & 8,058 \\
\hline
\multicolumn{6}{l}{\textit{TMDB Rating}} \\
\hline
M1 & Cast \% White & $-$0.06 & [$-$0.21, 0.09] & 0.459 & 8,622 \\
M1 & Crew \% White & 0.11 & [$-$0.12, 0.34] & 0.348 & 8,563 \\
M1 & Cast \% Female & $-$0.20$^{**}$ & [$-$0.34, $-$0.07] & 0.003 & 8,622 \\
M1 & Crew \% Female & 0.26$^{**}$ & [0.09, 0.42] & 0.002 & 8,563 \\
M2 & Cast \% White & $-$0.12 & [$-$0.27, 0.04] & 0.141 & 8,622 \\
M2 & Crew \% White & 0.07 & [$-$0.16, 0.31] & 0.540 & 8,563 \\
M2 & Cast \% Female & $-$0.20$^{**}$ & [$-$0.33, $-$0.06] & 0.005 & 8,622 \\
M2 & Crew \% Female & 0.11 & [$-$0.05, 0.28] & 0.177 & 8,563 \\
M3 & Cast \% White & $-$0.09 & [$-$0.26, 0.08] & 0.288 & 8,542 \\
M3 & Cast \% Female & $-$0.26$^{***}$ & [$-$0.40, $-$0.12] & $<$0.001 & 8,542 \\
M4 & Cast \% White & $-$0.15 & [$-$0.32, 0.02] & 0.082 & 8,542 \\
M4 & Cast \% Female & $-$0.23$^{**}$ & [$-$0.37, $-$0.09] & 0.001 & 8,542 \\
\hline
\end{tabular}
\end{table}

\clearpage
\section*{Supplementary Figures}
\label{figures}

\begin{figure}[htbp!]
    \centering
    \includegraphics[width=\linewidth]{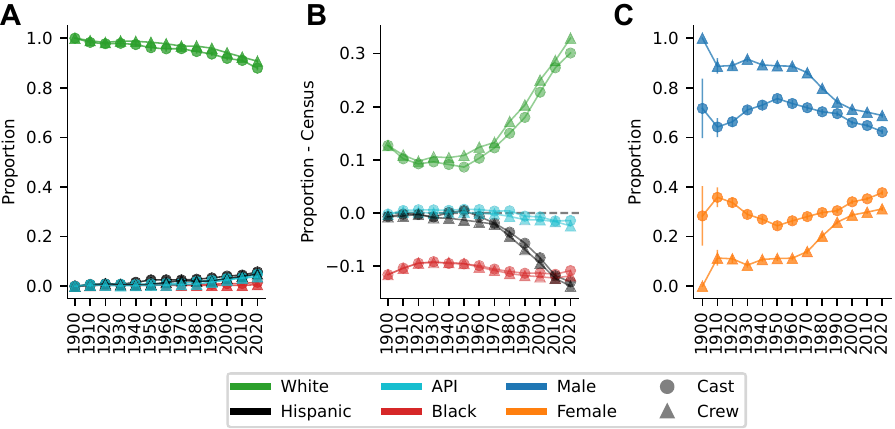}
    \caption{\textbf{Racial and gender representation of casts and crews in
    the top 5\% of films by popularity.} (\textbf{A}) Proportion of cast and crew positions occupied by each racial group over time. (\textbf{B}) Proportion of cast and crew positions occupied by each racial group relative to their U.S.\ Census population share per decade. (\textbf{C}) Proportion of cast and crew positions occupied by males and females over time.}
    \label{fig:si_top5_representation}
\end{figure}

\begin{figure}[htbp!]
    \centering
    \includegraphics[width=\linewidth]{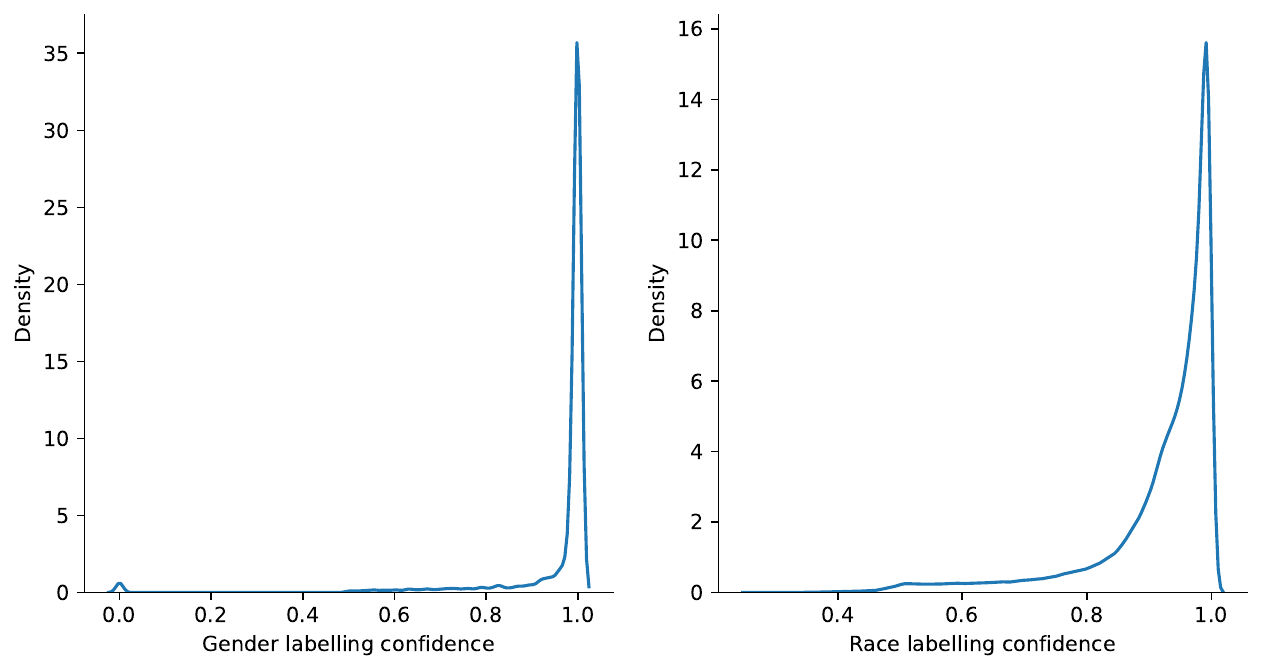}
    \caption{\textbf{Classifier confidence distributions for demographic inference.} Distribution of Genderize gender classification confidence \textbf{(left)} and NamePrism racial classification confidence \textbf{(right)} for cast and crew members in our dataset.}
    \label{fig:si_confidence_dist}
\end{figure}

\begin{figure}[htbp!]
    \centering
    \includegraphics[width=\linewidth]{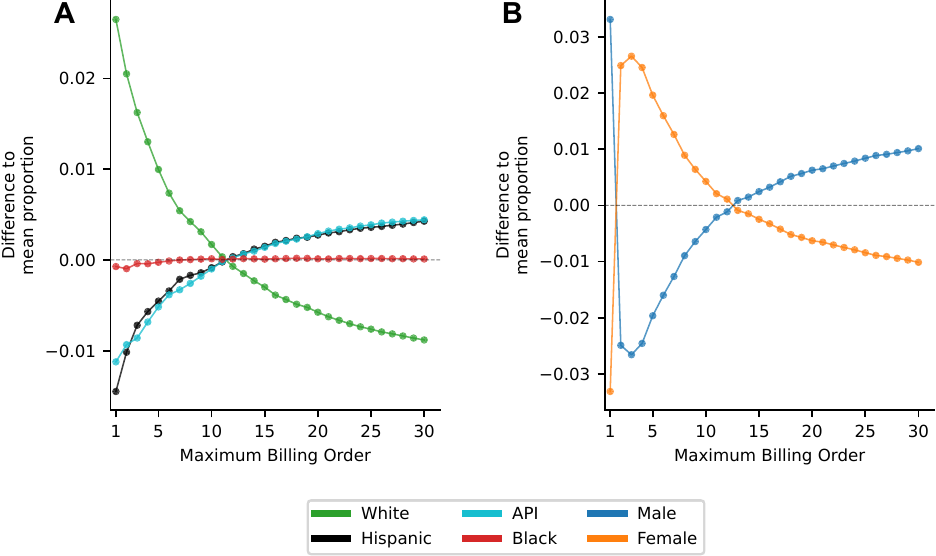}
    \caption{\textbf{Billing-order representation gaps in top-popularity films.} (\textbf{A}) Deviation between each racial group's representation at or above a given billing position and their overall representation, restricted to the top 5\% of films by popularity. (\textbf{B}) Equivalent analysis for gender groups.}
    \label{fig:si_billing_top5}
\end{figure}

\begin{figure}[h]
    \centering
    \includegraphics[width=0.7\textwidth]{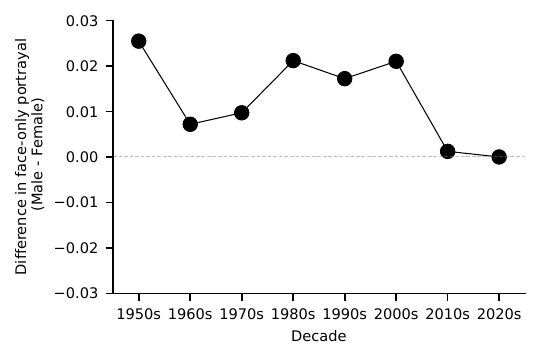}
    \caption{\textbf{Gender difference in face-only portrayal on U.S.\ movie posters by decade.} For each decade from the 1950s to the 2020s, points show the male--female difference in the proportion of depicted individuals classified as face only (rather than upper body or full body), computed from the poster sample of $11{,}951$ U.S.\ English-language films. Positive values indicate that men are more likely than women to be depicted face only; the dashed line marks parity (no gender difference). The difference is small throughout, peaking at approximately $2.5$ percentage points in the 1950s and narrowing to essentially zero by the 2010s and 2020s.}
    \label{fig:face_only_portrayal}
\end{figure}

\begin{figure}[htbp!]
    \centering
    \includegraphics[width=\linewidth]{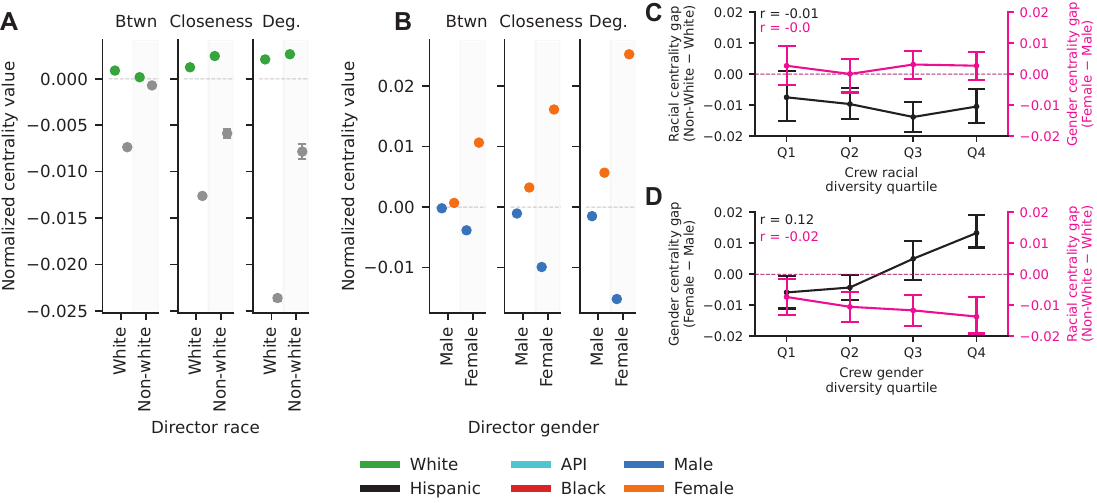}
    \caption{\textbf{Director demographics and actor narrative centrality.} (\textbf{A}) Normalized betweenness, closeness, and degree centrality for White and non-White actors in films directed by White and non-White directors, respectively. (\textbf{B}) Equivalent analysis for gender groups. (\textbf{C}) Association between crew racial diversity quartile and the cast racial centrality gap (Non-White minus White; black, left axis) and the cast gender centrality gap (Female minus Male; pink, right axis). (\textbf{D}) Association between crew gender diversity quartile and the cast gender centrality gap (Female minus Male; black, left axis) and the cast racial centrality gap (Non-White minus White; pink, right axis).}
    \label{fig:si_network_cross}
\end{figure}

\begin{figure}[htbp]
    \centering
    \includegraphics[width=\textwidth]{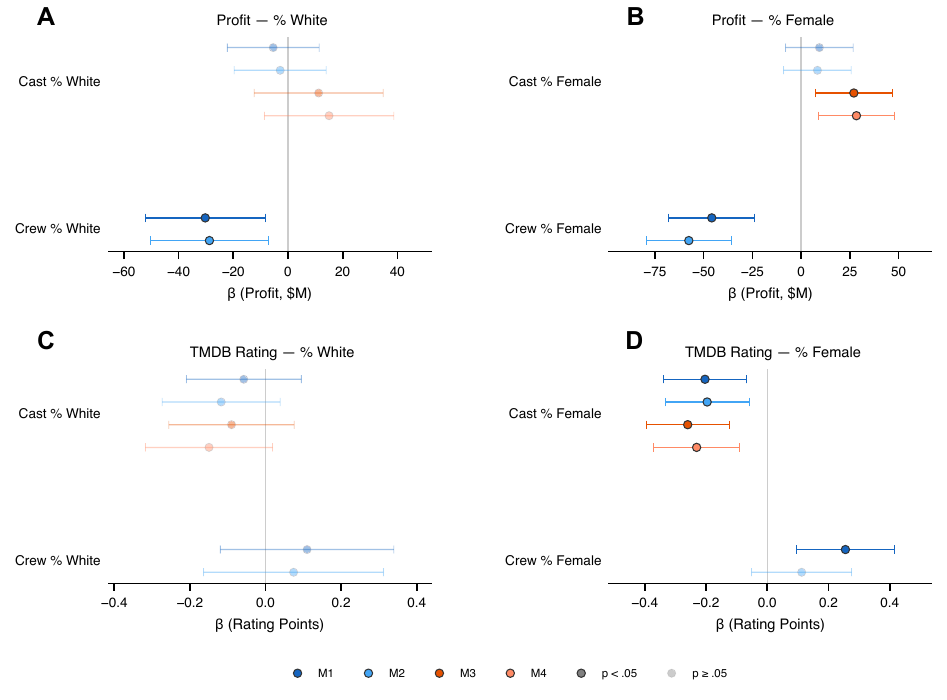}
    \caption{\textbf{Pooled OLS regression coefficients for all diversity variables across four model specifications.} Each panel shows the estimated coefficient ($\beta$) and 95\% confidence interval for a diversity variable on a given outcome across four models: M1 (decade fixed effects), M2 (year as a continuous control), M3 (decade fixed effects with crew diversity controlled), and M4 (year as a continuous control with crew diversity controlled). Panels are organized by outcome and predictor: (\textbf{A}) Cast \% White on profit; (\textbf{B}) Cast \% Female on profit; (\textbf{C}) Cast \% White on TMDB rating; (\textbf{D}) Cast \% Female on TMDB rating. M3 and M4 estimate cast variables only, as crew diversity enters these models as a control rather than a predictor. All models use HC3 robust standard errors and control for log budget, log runtime, log cast and crew size, and genre fixed effects. Darker points indicate $p < 0.05$; lighter points indicate $p \geq 0.05$.}
    \label{fig:si_pooled_full}
\end{figure}

\begin{figure}[htbp]
    \centering
    \includegraphics[width=\textwidth]{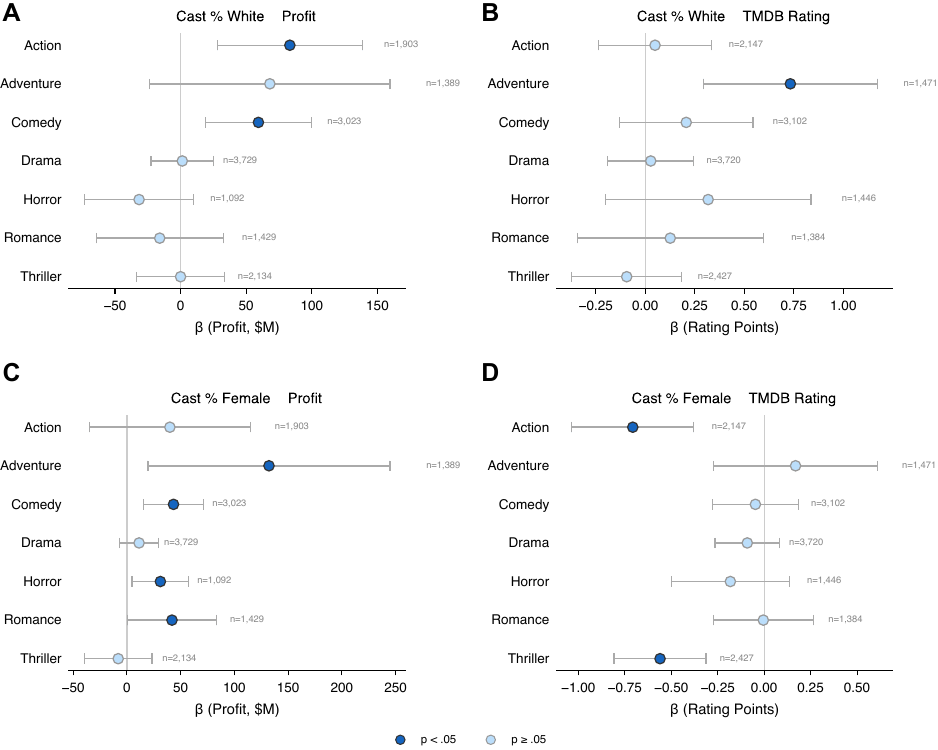}
    \caption{\textbf{Genre-stratified robustness check: within-genre regressions.} Pooled OLS coefficients ($\beta$) and 95\% confidence intervals for cast diversity on profit and audience ratings, estimated separately within each of the seven most frequent genre subsets: (\textbf{A}) Cast \% White on profit; (\textbf{B}) Cast \% White on TMDB rating; (\textbf{C}) Cast \% Female on profit; (\textbf{D}) Cast \% Female on TMDB rating. All models follow the M3 specification (decade fixed effects with crew \% White and crew \% Female as controls) and use HC3 robust standard errors. A film is assigned to a genre if it carries that genre tag regardless of other tags; a single film may therefore appear in multiple genre subsets. Sample sizes are annotated for each genre. Darker points indicate $p < 0.05$; lighter points indicate $p \geq 0.05$.}
    \label{fig:si_genre_within}
\end{figure}

\begin{figure}[htbp]
    \centering
    \includegraphics[width=\textwidth]{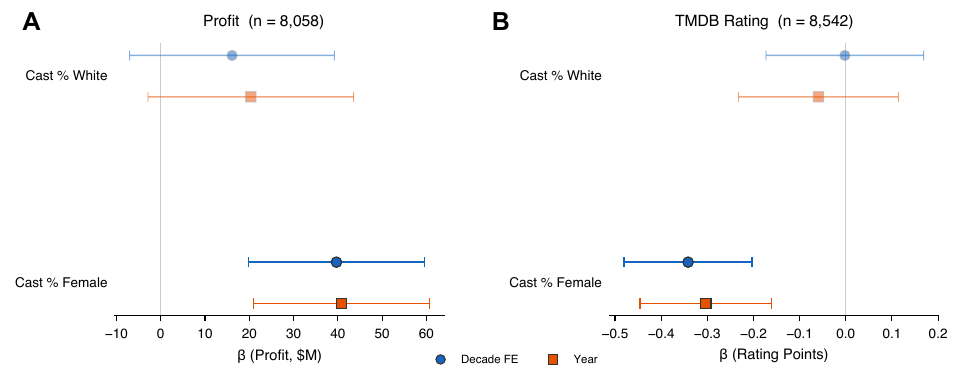}
    \caption{\textbf{Genre-stratified robustness check: multi-genre binary dummies on the full sample.} Pooled OLS coefficients ($\beta$) and 95\% confidence intervals for cast diversity on (\textbf{A}) profit and (\textbf{B}) TMDB rating, estimated on the full sample with binary indicator variables for each of the seven most frequent genres replacing the single primary-genre fixed effect. This specification accounts for films carrying multiple genre tags simultaneously. Two temporal specifications are shown: decade fixed effects (circles) and year as a continuous control (squares). All models include crew \% White and crew \% Female as controls and use HC3 robust standard errors.}
    \label{fig:si_genre_multidummy}
\end{figure}

\begin{figure}[htbp]
    \centering
    \includegraphics[width=0.8\textwidth]{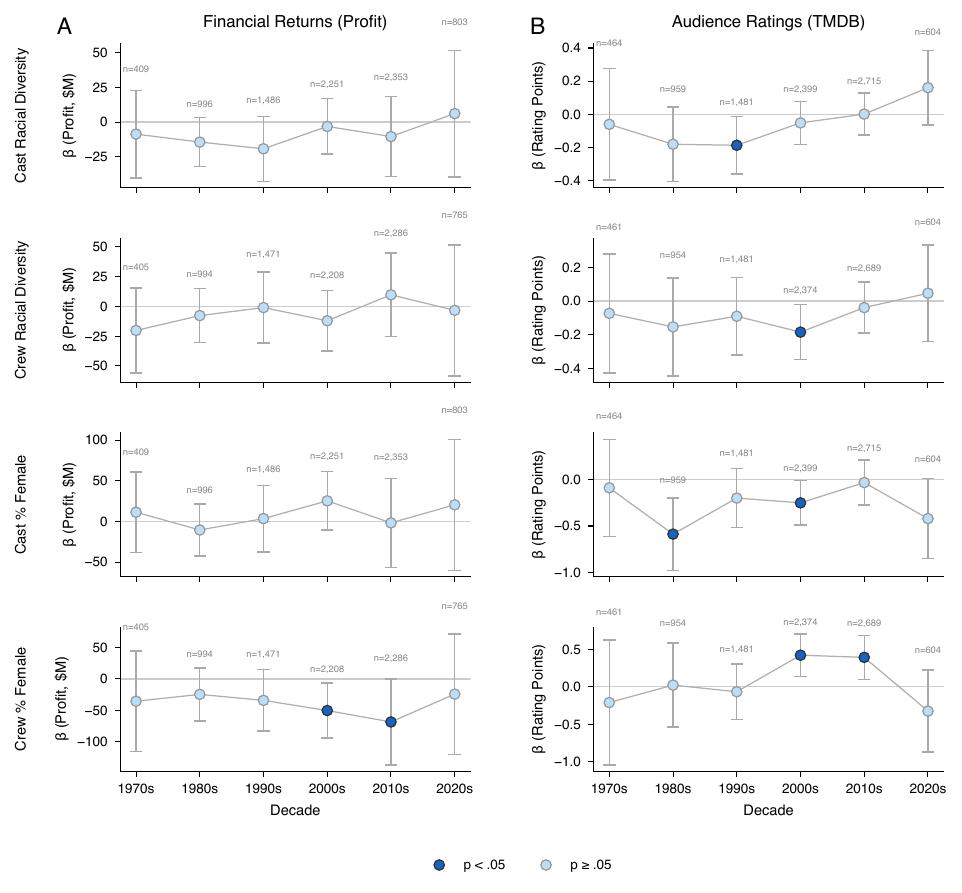}
    \caption{\textbf{Decade-by-decade replication: cast and crew diversity on financial returns and audience ratings.} Each panel plots the OLS coefficient ($\beta$) of a diversity variable, estimated separately within each decade from the 1970s to the 2020s, with 95\% confidence intervals and HC3 robust standard errors. (\textbf{A}) Financial returns, measured as profit (revenue minus budget, in millions of USD; full available sample $n = 9{,}516$). (\textbf{B}) Audience evaluations, measured as TMDB user rating (restricted to films with $\geq 50$ votes; full available sample $n = 19{,}133$). Rows correspond to four diversity variables: Cast \% White, Crew \% White, Cast \% Female, and Crew \% Female. All models control for log budget, log runtime, log cast size, log crew size, primary genre, and the within-dimension cross-diversity term. Darker points indicate $p < .05$; lighter points indicate $p \geq .05$. Per-decade sample sizes are annotated above each confidence interval.}
    \label{fig:si_decade_replication}
\end{figure}

\begin{figure}[htbp]
    \centering
    \includegraphics[width=\linewidth]{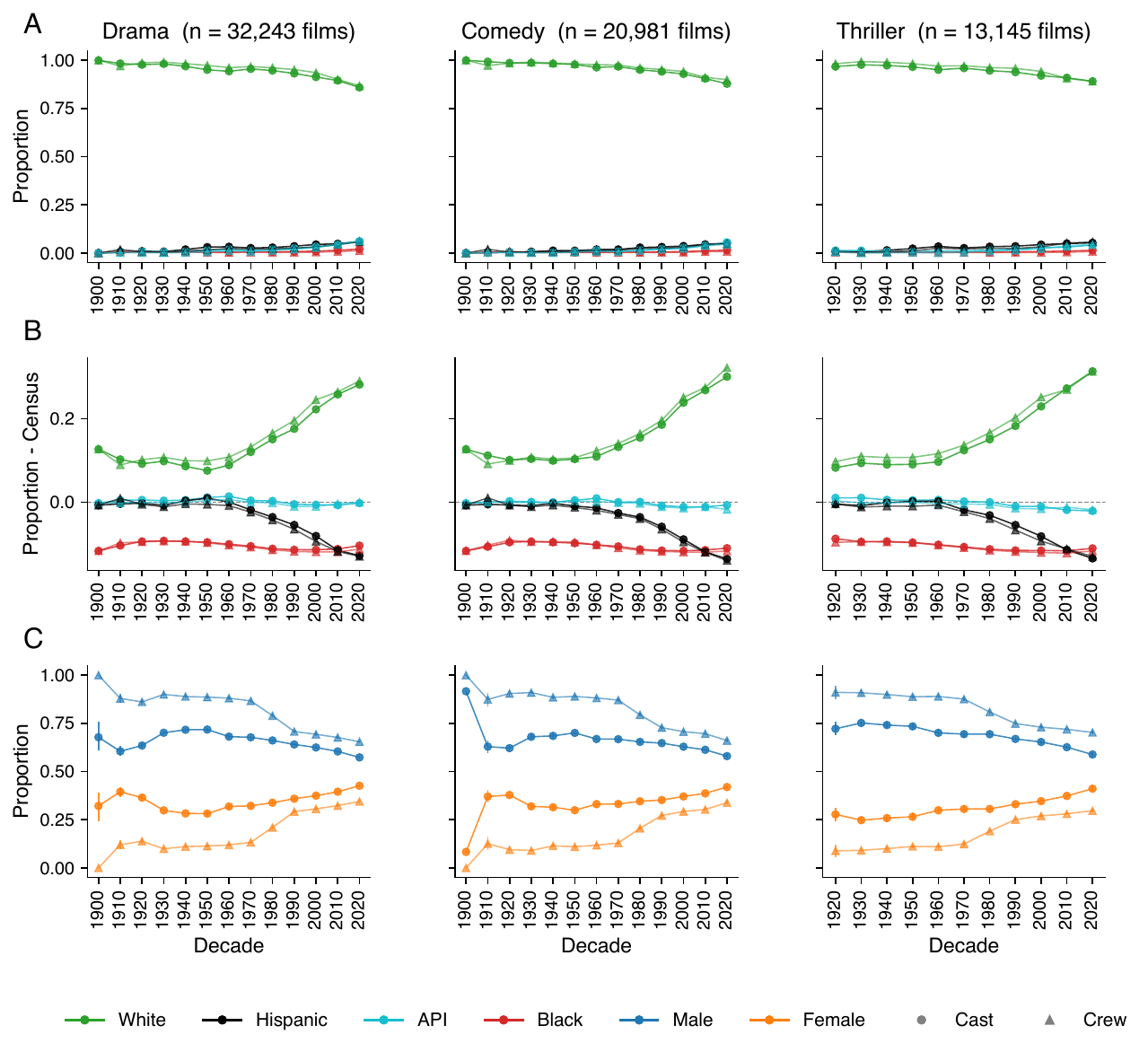}
    \caption{\textbf{Racial and gender representation within the three most frequent narrative genres.} Figure~1A-C replicated separately within Drama, Comedy, and Thriller (columns). Rows show (\textbf{A}) the proportion of cast (circles) and crew (triangles) positions held by each racial group by decade, (\textbf{B}) each group's representation relative to its U.S.\ Census population share, and (\textbf{C}) the male and female share of cast and crew positions over time. Genre is defined by TMDB tags and is multi-label, so a film contributes to every genre it carries; the number of films is given in each column header. White over-representation, the widening gap relative to the Census, and the persistent (though narrowing) gender gap each hold within all three genres.}
    \label{fig:si_genre_representation}
\end{figure}

\begin{figure}[htbp]
    \centering
    \includegraphics[width=\linewidth]{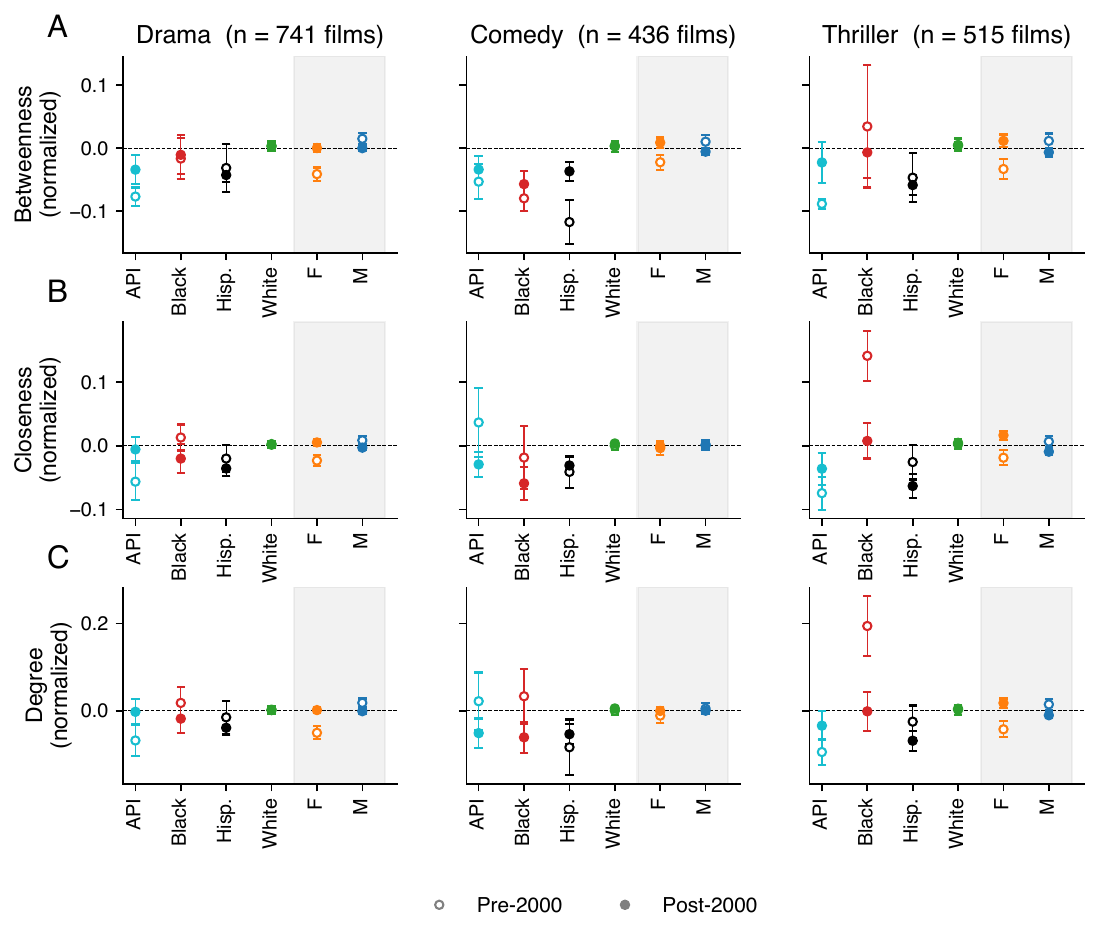}
    \caption{\textbf{Narrative centrality within the three most frequent narrative genres.} Figure~2C replicated within Drama, Comedy, and Thriller (columns) for normalized betweenness, closeness, and degree centrality (rows), with open markers for films released before 2000 and filled markers for films released after. Centrality is normalized within cast-size octile $\times$ release year, as in the main text. These panels use the Amazon X-Ray scene subset (per-genre film counts in each column header); because that subset is small, minority-group estimates split by era are underpowered: their wide whiskers reflect few observations (most visibly Black actors in Thriller), and these cells are shown for completeness. Racial minorities occupy less central positions than expected within each genre, consistent with the pooled result. Gender patterns also broadly replicate the pooled trend, with female actors shifting toward greater post-2000 centrality alongside a corresponding male decline, as in Figure~2C; as with the racial estimates, these per-genre, per-era gender values rest on a smaller sample than the pooled analysis and should be read as noisier accordingly.}
    \label{fig:si_genre_centrality}
\end{figure}

\begin{figure}[htbp]
    \centering
    \includegraphics[width=\linewidth]{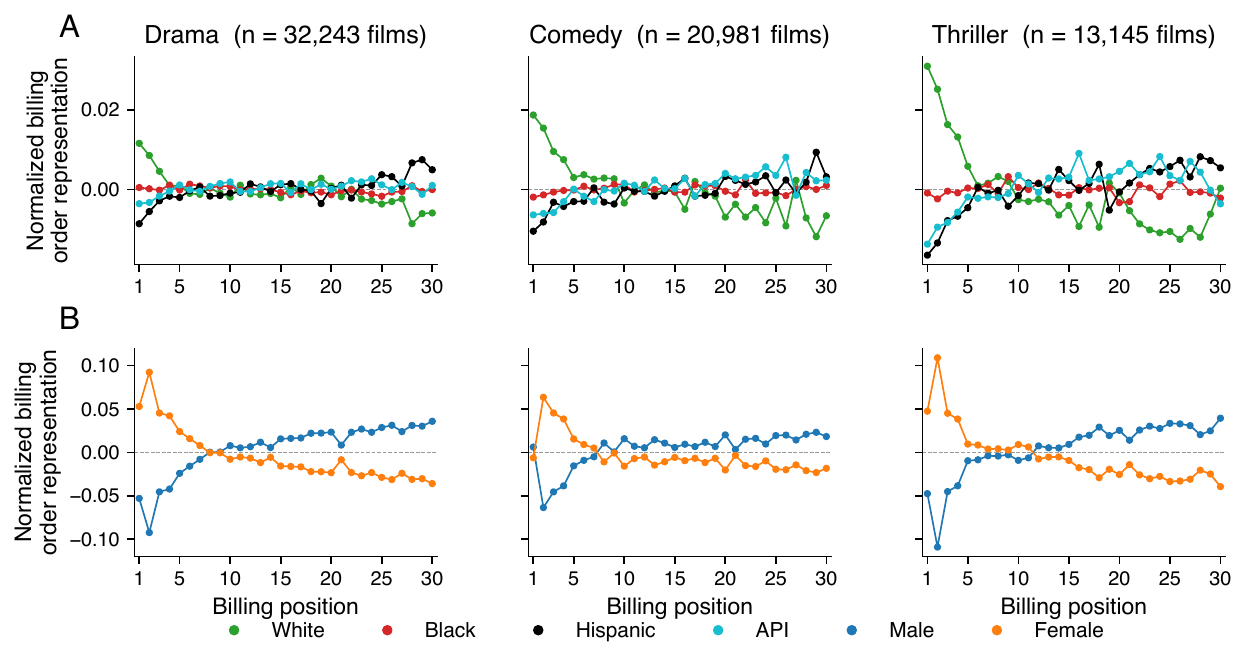}
    \caption{\textbf{Billing-position representation within the three most frequent narrative genres.} Figure~2D,E replicated within Drama, Comedy, and Thriller (columns) for racial groups (\textbf{A}) and gender (\textbf{B}). For each specific billing position (1--30), points show the observed share of that position held by a group minus its share under a null that permutes the billing-order column across the pooled cast (1{,}000 iterations); positive values indicate over-representation at that position. These panels use the full cast billing order (per-genre film counts in each column header), not the scene subset. Within every genre, White actors are over-represented at the top billing position and under-represented peripherally, while female actors are over-represented at the top-secondary positions (roughly 2--8) and under-represented beyond.}
    \label{fig:si_genre_billing}
\end{figure}

\begin{figure}[htbp]
    \centering
    \includegraphics[width=\linewidth]{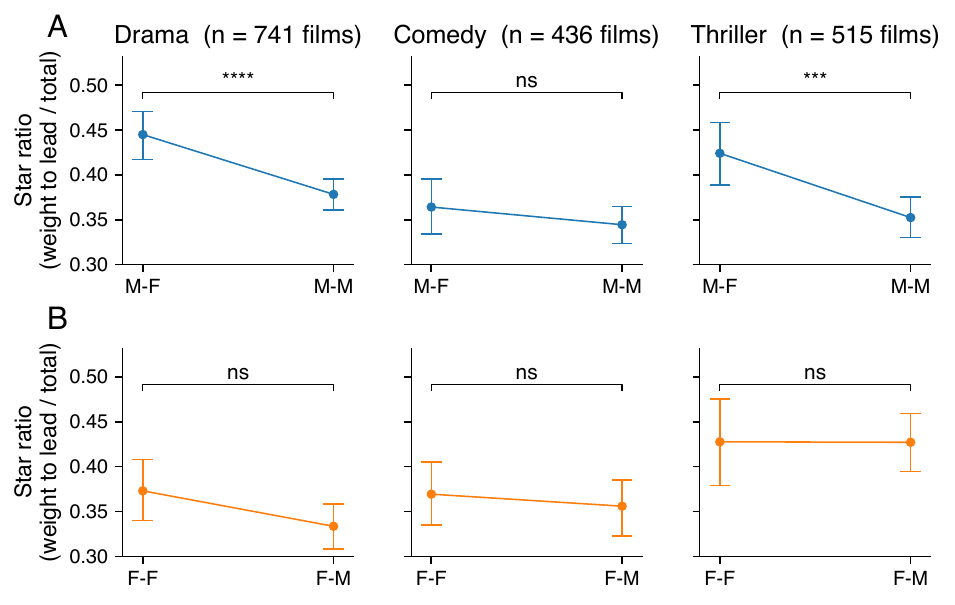}
    \caption{\textbf{Star ratio within the three most frequent narrative genres.} Figure~2F,G replicated within Drama, Comedy, and Thriller (columns) for male-led (\textbf{A}) and female-led (\textbf{B}) films. The star ratio is the share of a non-lead actor's co-appearance weight that is tied to the lead. Significance of the female-versus-male contrast is shown as a bracket with stars above each panel (two-sided $t$-test; \texttt{****}~$p \leq 10^{-4}$, \texttt{***}~$\leq 10^{-3}$, \texttt{**}~$\leq 10^{-2}$, \texttt{*}~$\leq 0.05$, else n.s.). These panels use the Amazon X-Ray scene subset (per-genre film counts in each column header); as with Supplementary Fig.~\ref{fig:si_genre_centrality}, these per-genre estimates rest on a smaller sample than the pooled analysis and should be read as noisier accordingly. In male-led films, female non-leads are more tethered to the lead than male non-leads; significant in Drama ($t = 4.1$) and Thriller ($t = 3.3$), not in Comedy ($t = 1.0$) though the direction is the same, possibly reflecting Comedy's smaller sample rather than a true absence of the effect. The asymmetry is absent in female-led films, consistent with the null pooled effect ($t \approx 0.99$).}
    \label{fig:si_genre_star_ratio}
\end{figure}

\begin{figure}[htbp]
    \centering
    \includegraphics[width=\linewidth]{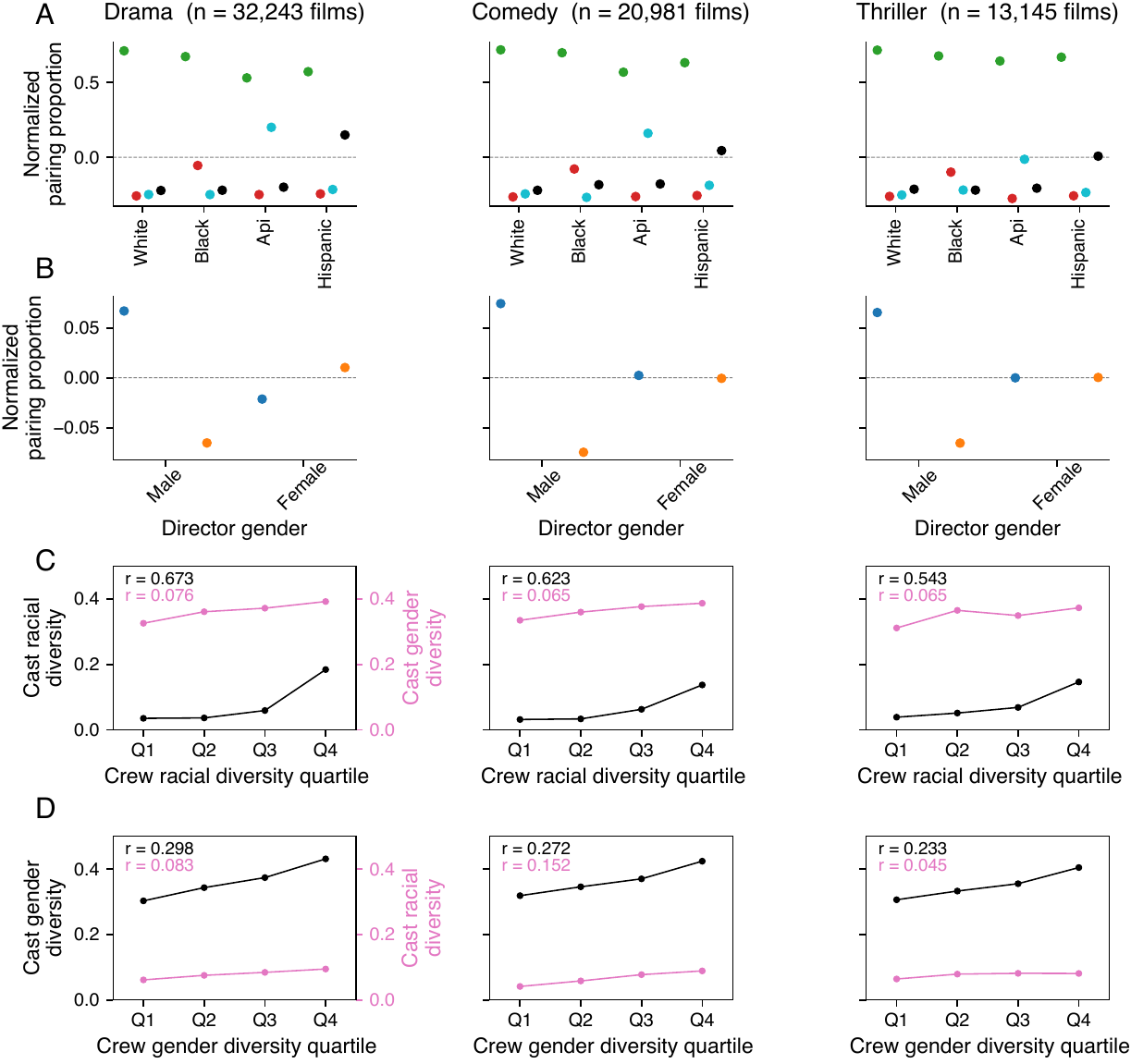}
    \caption{\textbf{Institutional gatekeeping within the three most frequent narrative genres.} Figure~4A-D replicated within Drama, Comedy, and Thriller (columns). Rows show (\textbf{A}) director--actor racial and (\textbf{B}) gender pairing relative to a bootstrapped null (top-three billing cut, 1{,}000 iterations), and the association between (\textbf{C}) crew racial-diversity quartile and cast racial diversity (black, left axis) and cast gender diversity (pink, right axis), and (\textbf{D}) crew gender-diversity quartile and cast gender diversity (black, left axis) and cast racial diversity (pink, right axis). In C and D both axes share a common 0--0.5 range, as in the main text, so the weakly correlated cross-diversity (pink) trend reads flat. Director--actor homophily and the within-dimension crew-to-cast diversity coupling replicate within each genre (crew-to-cast racial $r = 0.67$, $0.62$, $0.54$ and gender $r = 0.30$, $0.27$, $0.23$ for Drama, Comedy, Thriller); minority-director cells rest on few films within a genre.}
    \label{fig:si_genre_gatekeeping}
\end{figure}

\end{document}